\documentclass[aip,reprint,superscriptaddress]{revtex4-2}

\usepackage{gensymb}
\usepackage{notes2bib}
\usepackage{graphicx}
\usepackage{textcomp}
\usepackage{verbatim}
\usepackage{amsmath}
\usepackage{setspace}
\usepackage{multirow}

\usepackage[dvipsnames]{xcolor}

\usepackage[colorlinks=true,urlcolor=blue,linkcolor=blue,citecolor=blue,a4paper]{hyperref}

\begin{document}

\title{Combined Spectroscopy and Electrical Characterization of La:BaSnO$_\text{3}$\\ Thin Films and Heterostructures}
\author{Arnaud P. \surname{Nono Tchiomo}}
\affiliation{Department of Physics, University of Johannesburg, P.O. Box 524 Auckland Park 2006, Johannesburg, South Africa}
\affiliation{Van der Waals-Zeeman Institute, Institute of Physics, Science Park 904, 1098 XH Amsterdam, The Netherlands}
\author{Emanuela Carleschi}
\affiliation{Department of Physics, University of Johannesburg, P.O. Box 524 Auckland Park 2006, Johannesburg, South Africa}
\author{Aletta R. E. Prinsloo}
\affiliation{Department of Physics, University of Johannesburg, P.O. Box 524 Auckland Park 2006, Johannesburg, South Africa}
\author{Wilfried Sigle}
\affiliation{Max Planck Institute for Solid State Research, Heisenbergstr.\ 1, 70569 Stuttgart, Germany}
\author{Peter A. van Aken}
\affiliation{Max Planck Institute for Solid State Research, Heisenbergstr.\ 1, 70569 Stuttgart, Germany}
\author{Jochen Mannhart}
\affiliation{Max Planck Institute for Solid State Research, Heisenbergstr.\ 1, 70569 Stuttgart, Germany}
\author{Prosper Ngabonziza}
\email[corresponding author: ]{p.ngabonziza@fkf.mpg.de}
\affiliation{Department of Physics, University of Johannesburg, P.O. Box 524 Auckland Park 2006, Johannesburg, South Africa}
\affiliation{Max Planck Institute for Solid State Research, Heisenbergstr.\ 1, 70569 Stuttgart, Germany}
\author{Bryan P. Doyle}
\email[corresponding author: ]{bpdoyle@uj.ac.za}
\affiliation{Department of Physics, University of Johannesburg, P.O. Box 524 Auckland Park 2006, Johannesburg, South Africa}

\date{\today}

\begin{abstract}
For La-doped BaSnO$_\text{3}$ thin films grown by pulsed laser deposition, we combine chemical surface characterization and electronic transport studies to probe the evolution of electronic states in the band structure for different La-doping content. Systematic analyses of spectroscopic data based on fitting the core electron line shapes help to unravel the composition of the surface as well as the dynamics associated with increasing doping. This dynamics is observed with a more pronounced signature in the Sn 3d core level, which exhibits an increasing asymmetry to the high binding energy side of the peak with increasing electron density. The present results expand the current understanding of the interplay between the doping concentration, electronic band structure and transport properties of epitaxial La:BaSnO$_\text{3}$ films. 
\end{abstract}


\maketitle

The perovskite La-doped BaSnO$_\text{3}$ (La:BaSnO$_\text{3}$) is a novel transparent oxide semiconductor that exhibits outstanding room temperature (RT) electron mobility ($\mu_e$) with high carrier density together with a high optical transmittance~\cite{Kim2012b,Kim2012a,XLuo2012}. Owing to its unique electronic and optical properties, La:BaSnO$_\text{3}$ has the potential for applications in transparent electronics~\cite{Lee2017,SKim_2015,JYue_2018,ZWang_2019}, photovoltaics~\cite{Zhang2017,Shin_2013,Park2016,EFortunato_2007}, as well as in thermoelectric~\cite{JLi_2017,PRajasekaran_2020,TWu_2018,HCho_2019} and multifunctional perovskite-based optoelectronic devices~\cite{Ismail_2015,Krishnaswamy2016,Park2016}. Furthermore, its low-power consumption combined with its ability to be heavily doped and its good stability at high temperatures make La:BaSnO$_\text{3}$ a suitable material for integration in thermally  stable  capacitors, field effect transistors and power electronic devices~\cite{DGinley2000,XLuo2012,Prakash2017c,Krishnaswamy2016,Lee2017}.

The discovery of a RT $\mu_e$ of $320\text{ cm}^2\text{ V}^{-1}\text{s}^{-1}$ (with corresponding carrier density, $n = 8\times 10^{19} \text{ cm}^{-3}$) in La:BaSnO$_\text{3}$ single crystals~\cite{Kim2012b,Kim2012a,XLuo2012} stimulated intense investigation into this material~\cite{Lee2017}. Particularly, the potential of La:BaSnO$_\text{3}$ for device applications and heterostructures triggered considerable interest in thin films grown from this compound~\cite{Krishnaswamy2016,SKim_2015,UAlaan_2019,Sanchela2018,Cho2019,HCho_2019,Yu2016,Yoon2018,Niedermeier2017d,PVWadekar_2014,Nono-Tchiomo2019,FYFan_2018,Mizoguchi2013,KFujiwara_2017,ZWang_2019,JYue_2018,Raghavan2016c,Prakash2017c,Lebens-Higgins2016a,Paik2017c,ZWang_2019,HWang_2020,Postiglione2021,RZhang_2021,KGanguly_2017,HWang_2020,KGanguly_2015}. However, the reported $\mu_e$ in La:BaSnO$_\text{3}$ thin films have only reached a maximum value of $183\text{ cm}^2\text{ V}^{-1}\text{s}^{-1}$ ($n\simeq 1.2 \times 10^{20} \text{ cm}^{-3} $) for epitaxial films grown by molecular 
beam epitaxy (MBE)~\cite{Paik2017c}. Other growth techniques resulted in the following electron mobilities: $140\text{ cm}^2\text{ V}^{-1}\text{s}^{-1}$ ($n\simeq 5.2 \times 10^{20} \text{ cm}^{-3} $) for pulsed laser deposition (PLD)~\cite{Nono-Tchiomo2019}, $121\text{ cm}^2\text{ V}^{-1}\text{s}^{-1}$ ($n \simeq 4.0\times 10^{20} \text{ cm}^{-3} $) for high-pressure magnetron sputtering~\cite{RZhang_2021}, and $53\text{ cm}^2\text{ V}^{-1}\text{s}^{-1}$ ($n \simeq 2.0\times 10^{20} \text{ cm}^{-3} $) for chemical solution deposition~\cite{YHe_2021}. Various strategies to improve the mobility in La:BaSnO$_\text{3}$ epitaxial films have been explored. Such efforts include, for example, incorporation of undoped BaSnO$_\text{3}$ buffer layers to  compensate  for  the  lattice  mismatch  between the substrate and the active La:BaSnO$_\text{3}$ top layers~\cite{ZWang_2019,Prakash2017c,PVWadekar_2014,Paik2017c}, adsorption-controlled MBE for improved stoichiometry control~\cite{Paik2017c,APrakash_2017,Raghavan2016c,Lebens-Higgins2016a}, a very high-temperature grown insulating buffer layer to reduce the density of threading dislocations~\cite{Nono-Tchiomo2019}, and post growth annealing  processes~\cite{Yoon2018,Cho2019,WJLee_2016}. Besides the ongoing efforts for RT $\mu_e$ improvement, in order to gain a better understanding of the conduction mechanisms in La:BaSnO$_\text{3}$ films, it is important to establish a proper correlation between the transport characteristics and the behavior of the electronic states in the conduction band.  This is crucial because the high ambient $•\mu_e$ in La:BaSnO$_\text{3}$ has been proposed  to originate from both the small effective mass of the electrons at the conduction band minimum (CBM) \cite{Niedermeier2017d,Scanlon2013a}, which is associated with the largely dispersive Sn 5s conduction band, and the low optical phonon scattering rate~\cite{Prakash2017c,Sallis2013c}. 

\begin{figure*}[t]
	\centering 
	\includegraphics[width=1\textwidth]{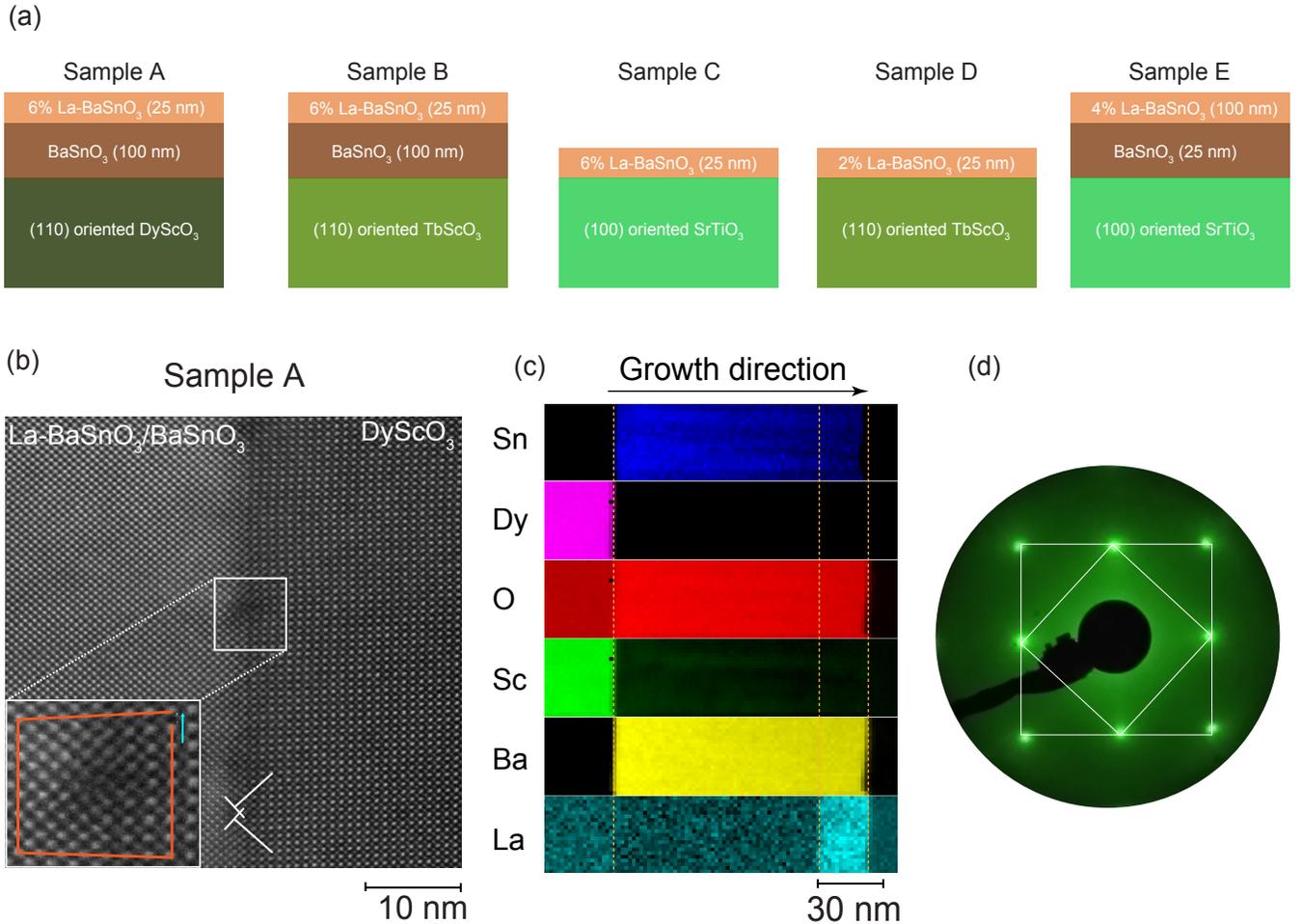}
	\caption{\label{Fig1}\small{(a) Schematic layout of the different thin film samples investigated in this study. (b) High-resolution  scanning  transmission  electron  microscopy  (HRSTEM)  image  of  a  representative  La:BaSnO$_3$/BaSnO$_3$ heterostructure (sample A). Misfit dislocations indicated by the $\perp$ symbols characterize the relaxed interface between the film and the $(110)$ oriented DyScO$_\text{3}$ substrate. The inset represents a high magnification around one of the misfits delimited by the white rectangle, where a lack of closure of the Burgers circuit is seen (orange rectangular-like contour). (c) Electron energy-loss spectroscopy (EELS) elemental mapping for the sample A showing the distribution of the elements in the sample.  The signal is divided into regions as indicated by the dashed orange lines.  The residual color in the La plot is noise. (d) A representative low energy electron diffraction (LEED) image taken at 48~eV displaying a clean La:BaSnO$_\text{3}$ (001) surface. The diffraction spots form square lattices (white rectangle drawn on the image) in reciprocal space.}}	 
\end{figure*}

Although several studies used photoemission spectroscopy techniques to investigate the electronic structure of La:BaSnO$_\text{3}$ films~\cite{Soltani2020,Joo2017b,Sallis2013c,Lebens-Higgins2016a,Lochocki2018b}, only a few reports have combined electronic transport and spectroscopic studies  to explore the evolution of electronic states in La:BaSnO$_\text{3}$ films and heterostructures at different La-doping levels~\cite{Sallis2013c,Lebens-Higgins2016a}. In particular, recent \textit{ex-situ} hard x-ray photoemission spectroscopy (HAXPES) experiments on La:BaSnO$_\text{3}$ films demonstrated that both the CBM and the valence band maximum (VBM), as well as the core electrons are effectively modified with increasing carrier density~\cite{Lebens-Higgins2016a}. Thus, this result calls for additional  combined spectroscopic and electrical characterizations to facilitate more quantitative exploration of the evolution of the intrinsic properties of La:BaSnO$_\text{3}$ films and heterostructures at different doping levels.

In the present paper we combine chemical surface analysis as a function of La doping using x-ray photoelectron spectroscopy (XPS) and electronic transport studies to explore the evolution of the electronic states in La:BaSnO$_\text{3}$ films and heterostructures. From the transport measurements, we extract the transport characteristics, as well as $n$ and $\mu_e$ of the La:BaSnO$_\text{3}$ samples. The surface properties of these samples are subsequently investigated using  spectroscopic techniques. 
A direct connection between the electronic transport characteristics and the spectroscopic data is demonstrated. We used XPS as a probing tool to measure the changes in the films spectra associated with the increasing amount of La$^{\text{3}+}$ dopant. Through the analysis and systematic fits of the core XPS spectra,  we are able to extract the binding energy values of the constituent elements along with the associated oxidation states. These data are consistent with the electron energy loss spectroscopy (EELS) data, as well as with the literature. By increasing the doping concentration, we observe shifts of the valence band leading edges toward higher binding energies, as well as increases in the states in the conduction band. More importantly, we provide a quantitative understanding of the effect of conduction band filling in La:BaSnO$_\text{3}$ films and heterostructures. This effect is manifested by  an increasing asymmetry in the line shape of the Sn~3d core spectra and leads to considering an additional plasmon satellite peak in the analysis of Sn~3d spectra. 

\begin{table*}[!t]
	\caption{\label{Tab_10}Electronic transport characteristics (carrier density and mobility) of the samples discussed in this study.}
	\begin{ruledtabular}
		\begin{tabular}{cccc}
			Sample name & Sample layout & Carrier density & Carrier mobility \\
			& & ($\times \text{10}^{\text{20}}$~cm$^{-\text{3}}$) & (cm$^\text{2}$~V$^{-\text{1}}$~s$^{-\text{1}}$) \\
			\hline
			A & 6\% La:BaSnO$_\text{3}$ (25~nm)/BaSnO$_\text{3}$ (100~nm)/DyScO$_\text{3}$ & $\text{1.24}\pm\text{0.02}$ & $\text{71}\pm\text{2}$ \\
			B & 6\% La:BaSnO$_\text{3}$ (25~nm)/BaSnO$_\text{3}$ (100~nm)/TbScO$_\text{3}$ & $\text{4.92}\pm\text{0.05}$ & $\text{20}\pm\text{1}$ \\
			C & 6\% La:BaSnO$_\text{3}$ (25~nm)/SrTiO$_\text{3}$ & $\text{4.05}\pm\text{0.05}$ & $\text{91}\pm\text{2}$ \\
			D & 2\% La:BaSnO$_\text{3}$ (25~nm)/TbScO$_\text{3}$ & $\text{1.35}\pm\text{0.02}$ & $\text{75}\pm\text{2}$ \\
			E & 4\% La:BaSnO$_\text{3}$ (100~nm)/BaSnO$_\text{3}$ (25~nm)/SrTiO$_\text{3}$ & $\text{0.80}\pm\text{0.05}$ & $\text{18}\pm\text{1}$ \\
			
		\end{tabular}
	\end{ruledtabular} 
\end{table*}

Epitaxial La:BaSnO$_\text{3}$ films and heterostructures (samples labeled A to E in Table~\ref{Tab_10}) were prepared by PLD ($\lambda = 248$ nm). Prior to deposition, the (100) oriented SrTiO$_\text{3}$, (110) oriented  DyScO$_\text{3}$ and TbScO$_\text{3}$ crystalline  substrates  $(5\times5\times1 \text{ mm}^3)$ were terminated in situ at 1300\degree C using a CO$_2$ laser substrate heating system~\cite{WBraun_2020}. Figure~\ref{Fig1}\textcolor{blue}{(a)} depicts a schematic view of the sample types investigated. The films were grown from La:BaSnO$_3$ targets of 2\%, 4\% and 6\% La doping contents. Details about the growth and systematic characterization of the films are provided in Ref.~\cite{Nono-Tchiomo2019} [see details on the PLD growth conditions in Table \textcolor{blue}{S1} of the supplemental information].
 \begin{figure}[!t]
	\centering 
	\includegraphics[width=0.475\textwidth]{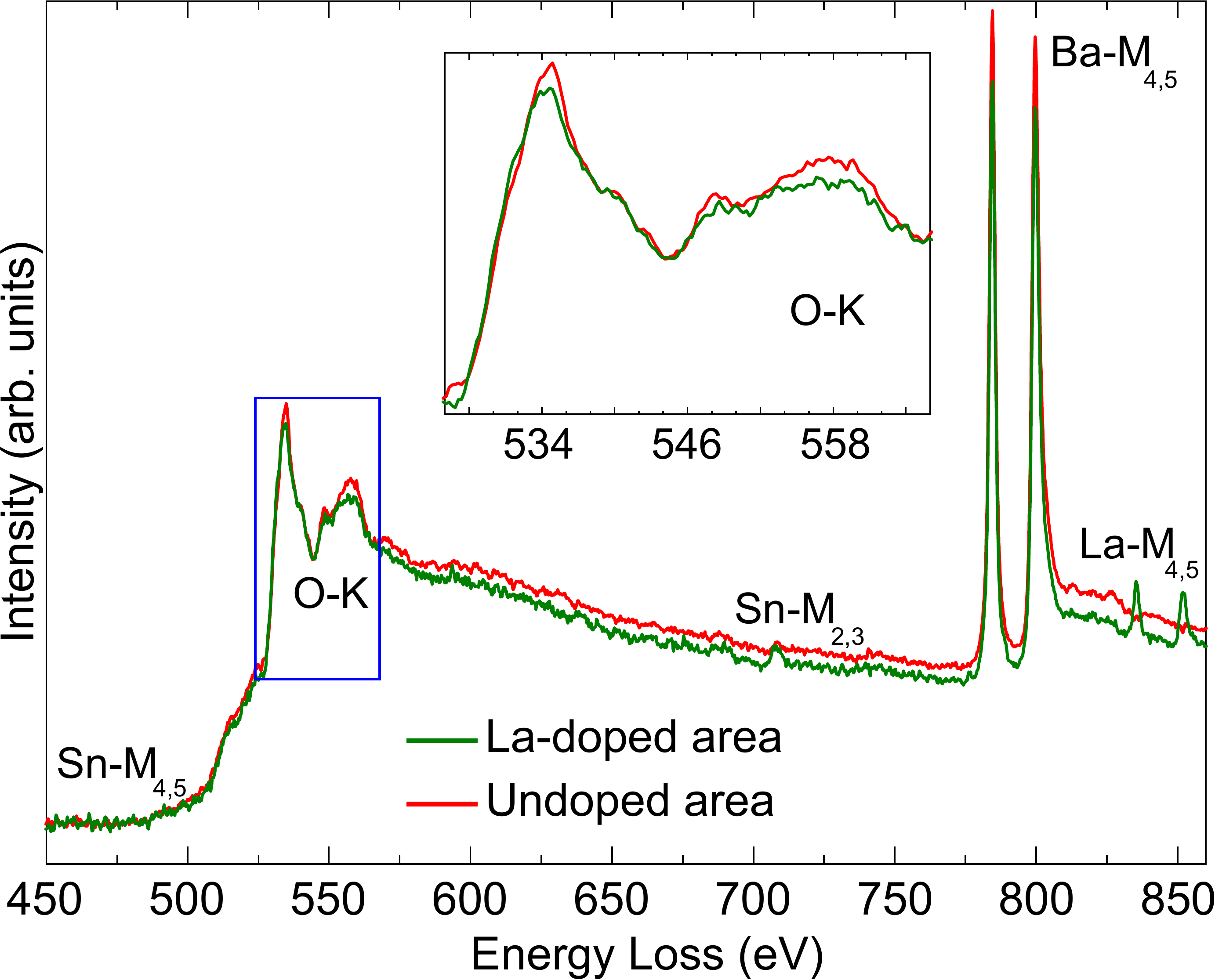}
	\caption{\label{Fig2}\small{Electron energy loss spectra of  a representative La:BaSnO$_3$/BaSnO$_3$ heterostructure. Two regions in the specimen were investigated: an area with only  BaSnO$_\text{3}$ (red curve) and another with only La:BaSnO$_\text{3}$ (green curve). The ionized edges of La-M$_{\text{4,5}}$, Ba-M$_{\text{4,5}}$ and Sn-M$_{\text{4,5}}$ (corresponding to the excitation of 3d states), Sn-M$_{\text{2,3}}$ (corresponding to the excitation of 3p states), and O-K (corresponding to the excitation of the 1s state) are resolved. The inset shows the enlargement of the O-K edge delimited by the blue rectangle.}}	 
\end{figure}

Electrical  transport  properties  were measured in a physical property measurement system (PPMS)  in  a  van  der  Pauw  geometry obtained by  wire  bonding  aluminum wires  to  the  samples'  corners [see Fig.~\textcolor{blue}{S1(a)} of the supplementary  information].  The carrier concentration, $n$, and   the electron mobility, $\mu_e$, were determined following the procedure discussed elsewhere~\cite{Nono-Tchiomo2019,PNgabonziza_2016,PNgabonziza_2018,PNgabonziza_2022}. 
Table~\ref{Tab_10} gives the carrier density and electron mobility of the different sample types investigated in this study. The La:BaSnO$_\text{3}$ samples analyzed here in both spectroscopy and transport experiments have a RT carrier concentration ranging from   $n=0.80\times 10^{20}$ to $4.92\times 10^{20}\text{ cm}^{-3} $; and carrier mobility from $\mu_e=18 $ to $91\text{ cm}^{2} \text{ V}^{-1}\text{ s}^{1} $. The measured carrier densities reported are smaller than the expected values associated with the nominal concentration of the La dopant. In fact, the doping levels corresponding to  2\%, 4\% and 6\% La are $\text{2.72}\times \text{10}^{\text{20}}$~cm$^{-\text{3}}$, $\text{5.44}\times \text{10}^{\text{20}}$~cm$^{-\text{3}}$ and $\text{8.16}\times \text{10}^{\text{20}}$~cm$^{-\text{3}}$, respectively. The observed discrepancy can be ascribed to the high density of dislocation defects present in the films, as seen in the weak beam dark field scanning transmission electron microscopy micrographs [Fig.~\textcolor{blue}{S1(b)-(d)}]. These defects arise because the films are grown on substrates to which they are poorly lattice matched; and this often results in the reduction of the carrier density and electron mobility in epitaxial La:BaSnO$_3$ films~\cite{Nono-Tchiomo2019,Paik2017c,Raghavan2016c,Kim2012a} [see more details in section \textcolor{blue}{S1} of the supplemental information].

Following the procedure described in Ref.~\cite{NonoTchiomo2018}, the samples were systematically cleaned in ultra-high vacuum (UHV) before photoemission experiments. The surface structure of the samples was characterized using low-energy electron diffraction (LEED) [see Fig.~\textcolor{blue}{S2(a)}-\textcolor{blue}{(b)} in supplementary information]. The surface of the clean La:BaSnO$_3$ $(001)$ surface showed a stable 1$\times$1 surface structure [Fig.~\ref{Fig1}\textcolor{blue}{(d)}].  The diffraction spots form square lattices in reciprocal space corresponding to the cubic lattice structure of BaSnO$_\text{3}$ in real space, thus indicating the high crystallinity of the films \cite{Hove1979}. The cleanliness of the samples was checked by monitoring the LEED patterns directly after an annealing cycle, and also by comparing the XPS survey scans before and after cleaning [see Fig.~\textcolor{blue}{S2} in supplementary information]. 
\begin{figure}[!t]
	\centering 
	\includegraphics[width=0.45\textwidth]{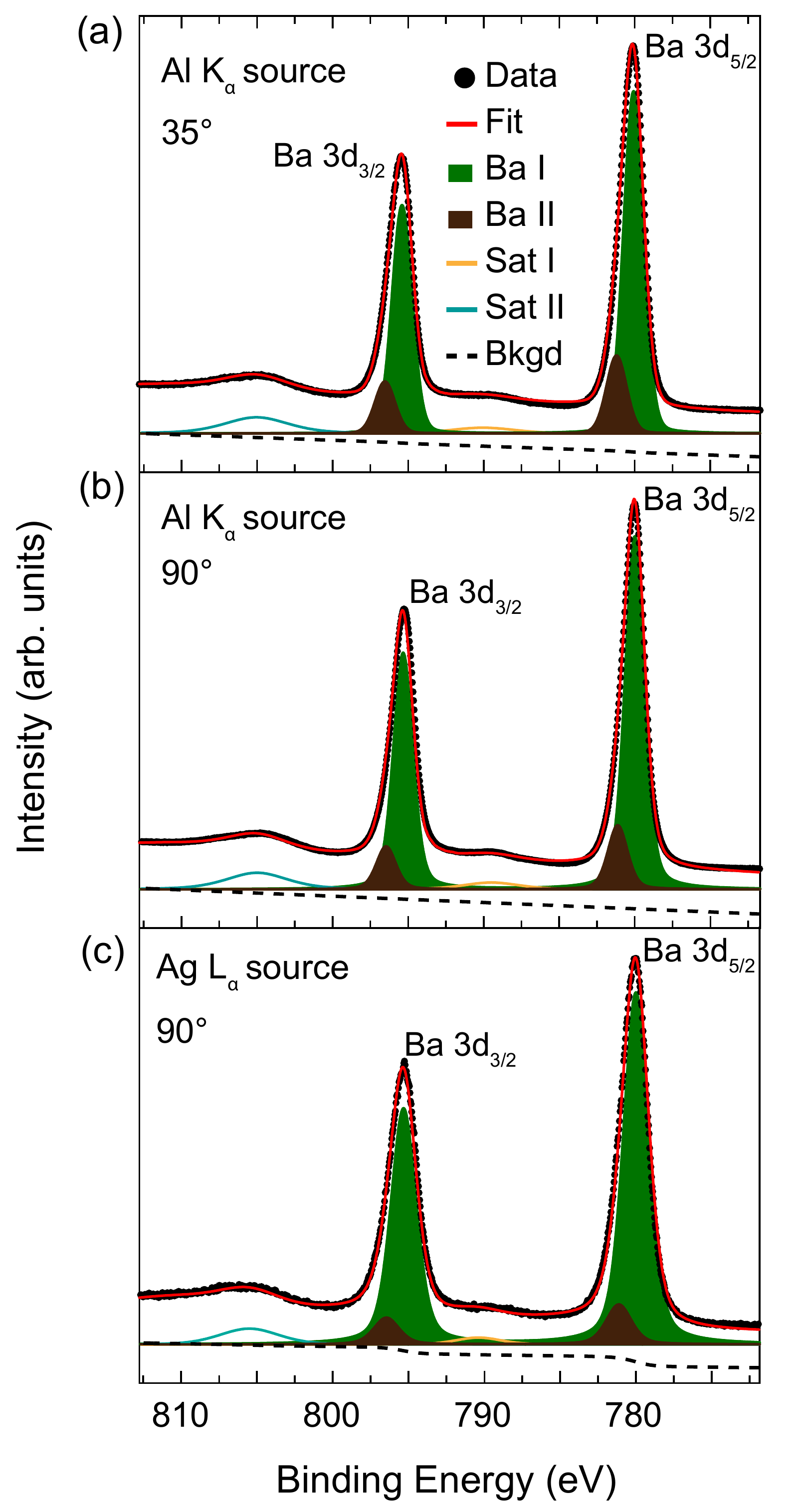}
	\caption{\label{Fig3}\small{Angle- and energy-dependent XPS spectra of the Ba~3d core level for the sample B. After subtracting a Shirley background (black dashed lineshape), the Ba~3d core lines are fitted (red lineshape) with two Voigt doublet peaks. The spectra in (a) and (b) were acquired with an Al~K$_{\alpha}$ anode (Al~K$_{\alpha}$ source) at 35$\degree$ and 90$\degree$ take-off angles, respectively. The spectrum in (c) was measured with an Ag~L$_\alpha$ anode (Ag~L$_\alpha$ source) at 90$\degree$ take-off angle.  Two satellite features (cyan and orange peaks) are resolved in the spectra. The peak heights
were normalized for clarity.}}	 
\end{figure}
 
Figure~\ref{Fig1}\textcolor{blue}{(b)} shows a representative scanning transmission electron microscopy (STEM) image of the investigated La:BaSnO$_\text{3}$  heterostructure (sample A). To investigate the stoichiometry and evaluate the composition  of the epitaxial layers, EELS measurements were performed  on the cross section of the samples during STEM characterization. Figure~\ref{Fig1}\textcolor{blue}{(c)} displays a representative EELS elemental mapping (atomic layer distribution) of the sample A for a total layer thickness of 120~nm of which the La-doped layer is 25~nm thick. The black color suggests a zero intensity, whereas the other colors indicate the presence of different elements (La, Ba, Sn, Dy, Sc, O) that are resolved in the scanned region. The elemental composition of the different layers of the La:BaSnO$_\text{3}$/BaSnO$_\text{3}$ heterostructure is also evidenced in the EELS spectra for the same sample presented in Fig.~\ref{Fig2}. Looking at all the ionized edges, it can be seen that the spectra of the BaSnO$_\text{3}$ (red curve) and La:BaSnO$_\text{3}$ (green curve) are very similar, apart from the transfer of spectral weight to the La peaks in the latter. This spectral weight transfer is illustrated by the reduction of the intensity at the Ba-M edge. Intensity reduction is also noticeable at the O-K edge, and could well be due to electronic (charge) modulation associated with the substitution of the bivalent Ba$^{2+}$ atoms with the trivalent La$^{3+}$ ones \cite{liu2021recent}. Furthermore,  the line shape of the spectra around the O-K feature [see inset Fig.~\ref{Fig2}] indicate good stoichiometry (no oxygen deficiency) in bulk layers, since these are in good agreement with the O-K EELS spectra reported for stoichiometric BaSnO$_\text{3}$ and La:BaSnO$_\text{3}$ films \cite{Paik2017c,Wang2015c,yun2018electronic}.
\begin{table}[!t]
	\caption{\label{Tab_20}Surface atomic percentages for the samples measured in XPS.}
	\setlength{\tabcolsep}{12pt}

		\begin{tabular}{cccc}
		\hline
		\hline
			Sample name &  Ba & Sn & O \\
		\hline
			B &  20\% & 16\% & 64\% \\
			C &  25\% & 15\% & 60\% \\
			D &  22\% & 16\% & 62\%\\
			E &  23\% & 13\% & 64\% \\
		\hline
		\hline	
		\end{tabular}
\end{table}

Table~\ref{Tab_20} presents the elemental composition of the surface of the films obtained from the fit of the XPS spectra. The composition is consistent throughout the surface of the various samples, with a larger proportion of Ba compared to Sn. Most importantly, the oxygen proportion in these high temperature annealed samples demonstrates the stability of the oxygen atoms in La:BaSnO$_\text{3}$ \cite{Kim2012b,XLuo2012}, and highlights the robustness of the sample cleaning procedure.
\begin{table}
	\caption{\label{Tab1}\small{Peak ratios for angle-dependent XPS of the Ba~3d spectra.}}
	\begin{ruledtabular}
		\begin{tabular}{ccccc}
			& \multicolumn{1}{c}{\small{Peak}} & \multicolumn{1}{c}{\small{Peak}} & \multicolumn{1}{c}{\small{Relative}} & \multicolumn{1}{c}{\small{FWHM}} \\
			& \multicolumn{1}{c}{\small{assignment}} & \multicolumn{1}{c}{\small{position}} & \multicolumn{1}{c}{\small{intensity}} & \multicolumn{1}{c}{\small{}} \\
			& \multicolumn{1}{l}{\small{}}	& \multicolumn{1}{c}{\small{($\pm\text{0.05~eV}$)}} & \multicolumn{1}{c}{\small{($\pm\text{1}~\%$)}} & \multicolumn{1}{c}{\small{($\pm\text{0.05~eV}$)}}  \\
			\hline
			\multicolumn{1}{c}{\multirow{1}{*}{\small{35$\degree$}}} & \multicolumn{1}{c}{\small{Ba I}} & \small{780.08} & \small{82} & \multicolumn{1}{c}{\small{1.58}} \\
			\multicolumn{1}{c}{\multirow{1}{*}{\small{Al anode}}} & \multicolumn{1}{c}{\small{Ba II}} & \small{781.21} & \small{18} & \multicolumn{1}{c}{\small{1.58}} \\
			& & & &\\
			\multicolumn{1}{c}{\multirow{1}{*}{\small{90$\degree$}}} & \multicolumn{1}{c}{\small{Ba I}} & \small{780.00} & \small{85} & \multicolumn{1}{c}{\small{1.53}} \\
			\multicolumn{1}{c}{\multirow{1}{*}{\small{Al anode}}} & \multicolumn{1}{c}{\small{Ba II}} & \small{781.13} & \small{15} & \multicolumn{1}{c}{\small{1.53}} \\
			& & & &\\
			\multicolumn{1}{c}{\multirow{1}{*}{\small{90$\degree$}}} & \multicolumn{1}{c}{\small{Ba I}} & \small{780.00} & \small{90} & \multicolumn{1}{c}{\small{1.97}} \\
			\multicolumn{1}{c}{\multirow{1}{*}{\small{Ag anode}}} & \multicolumn{1}{c}{\small{Ba II}} & \small{781.13} & \small{10} & \multicolumn{1}{c}{\small{1.97}} \\
		\end{tabular}
	\end{ruledtabular} 
\end{table}  

\begin{figure}[!t]
	\centering 
	\includegraphics[width=0.45\textwidth]{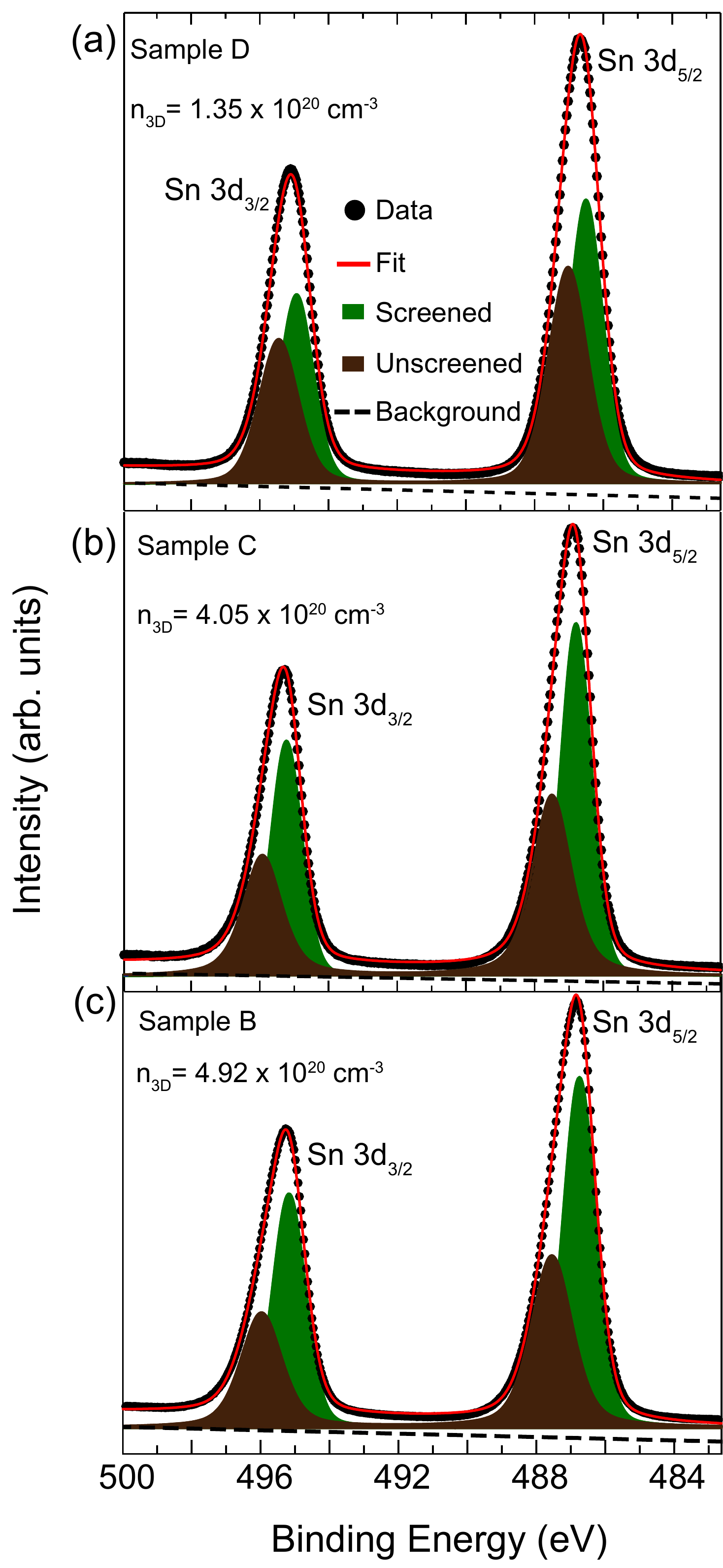}
	\caption{\label{Fig4}\small{XPS spectra around the Sn~3d regions for three samples of different total carrier density: (a) Sample D with $\text{n}_{_{3D}}= 1.35\times 10^{20}$ cm$^{-3}$, (b) Sample C with $\text{n}_{_{3D}}= 4.05\times 10^{20}$ cm$^{-3}$, and (c) Sample B with $\text{n}_{_{3D}}= 4.92\times 10^{20}$ cm$^{-3}$. After subtracting a Shirley background (black dashed lineshape), the  Sn~3d spectra are fitted (red lineshape) with two Voigt doublet components. The data were acquired at normal emission with the Al~K$_{\alpha}$ excitation source.}}	 
\end{figure}

\begin{table*}[!t]
	\caption{\label{Tab2}Fitted parameters of the Sn~3d regions along with the calculated carrier density in each sample.}
	\begin{ruledtabular}
		\begin{tabular}{ccccccc}
			Sample name &  \multicolumn{1}{c}{Peak assignment} & \multicolumn{1}{c}{Peak position} &  \multicolumn{1}{c}{FWHM}  &  \multicolumn{1}{c}{Relative intensity} &  \multicolumn{1}{c}{Satellite energy} & \multicolumn{1}{c}{Carrier density} \\
			& 	& \multicolumn{1}{c}{($\pm\text{0.05~eV}$)} & \multicolumn{1}{c}{($\pm\text{0.05}$)} & \multicolumn{1}{c}{($\pm\text{1}\%$)} & \multicolumn{1}{c}{(eV)} & \multicolumn{1}{c}{($\times \text{10}^{\text{20}}$~cm$^{-\text{3}}$)}   \\
			\hline
			\multicolumn{1}{c}{\multirow{2}{*}{E}} & \multicolumn{1}{c}{\small{Screened}} & \small{486.76} & 1.07 & \multicolumn{1}{c}{54} & \multicolumn{1}{c}{\multirow{2}{*}{0.30}} & \multicolumn{1}{c}{\multirow{2}{*}{$\text{0.8}\pm\text{0.05}$}}  \\
			& \multicolumn{1}{c}{\small{Unscreened}} & 487.06 & 1.66 & 46 &  &   \\
			
			\multicolumn{1}{c}{\multirow{2}{*}{D}} & \multicolumn{1}{c}{\small{Screened}} & \small{486.51} & 1.15 & \multicolumn{1}{c}{52} & \multicolumn{1}{c}{\multirow{2}{*}{0.52}} & \multicolumn{1}{c}{\multirow{2}{*}{$\text{1.35}\pm\text{0.02}$}}  \\
			& \multicolumn{1}{c}{\small{Unscreened}} & 487.03 & 1.34 & 48 &  &   \\
			\multicolumn{1}{c}{\multirow{2}{*}{C}} & \multicolumn{1}{c}{\small{Screened}} & 486.80 &  1.06 & 57 & \multicolumn{1}{c}{\multirow{2}{*}{0.70}} & \multicolumn{1}{c}{\multirow{2}{*}{$\text{4.05}\pm\text{0.05}$}}  \\
			& \multicolumn{1}{c}{\small{Unscreened}} & 487.50 & 1.34 & 43 &  &  \\
			\multicolumn{1}{c}{\multirow{2}{*}{B}} & \multicolumn{1}{c}{\small{Screened}} & 486.75 & 1.12 & 60 & \multicolumn{1}{c}{\multirow{2}{*}{0.80}} & \multicolumn{1}{c}{\multirow{2}{*}{$\text{4.92}\pm\text{0.05}$}}  \\
			& \multicolumn{1}{c}{\small{Unscreened}} & 487.55 & 1.40 & 40 &  &  \\
		\end{tabular}
	\end{ruledtabular}  
\end{table*}
Figure~\ref{Fig3} represents the XPS spectra of the Ba~3d core electrons together with the Voigt function fits. The Ba~3d  spectra of all the samples showed an asymmetric line shape, suggesting the presence of multiple components in the core level. Two symmetric Voigt doublets (Ba~I and Ba~II) were used to fit the spectra. The main doublet, Ba~I, located at the binding energy of 780.00~eV, is assigned to lattice barium in the  Ba$^{\text{2}+}$ state,  consistent with previous spectroscopic results on powder and epitaxial thin films of BaSnO$_\text{3}$ \cite{Larramona1989a,Jaim2017a}. 
The second doublet, Ba~II, situated at higher binding energy (781.13~eV), has been attributed to a surface character in several reports on epitaxial BaTiO$_\text{3}$ films \cite{Rault2013,Li2005,Li2008}. This component was suggested to originate either from under-coordinated barium at a BaO  terminated surface, or from lattice relaxation \cite{Li2005,Li2008}. 

In comparing the XPS spectra for cleaned surfaces with those of  surfaces measured as-inserted [see Figs.~\textcolor{blue}{S3} and \textcolor{blue}{S4} of the supplemental information], the relative intensity of  Ba~II was observed to decrease after the treatment of the surfaces, while that of Ba~I increases. This trend suggests that Ba~II could be a surface component, which is amplified with contamination. The nature of the Ba~II peak was carefully investigated by carrying out a systematic analysis of its fraction with respect to the probing depth. This was achieved by performing angle-dependent XPS measurements as depicted in Fig.~\ref{Fig3}. The measurements were first performed using the Al~K$_{\alpha}$ anode (photon energy of 1486.71~eV) at electron take-off angles of 35$\degree$ [Fig.~\ref{Fig3}\textcolor{blue}{(a)}] and  90$\degree$ (normal emission) [Fig.~\ref{Fig3}\textcolor{blue}{(b)}], and later the excitation source was changed to Ag~L$_\alpha$ anode (photon energy of 2984.31~eV) for acquisition at normal emission [Fig.~\ref{Fig3}\textcolor{blue}{(c)}]. For the take-off angle of 35$\degree$, the photoelectrons emitted originate from a region nearer the surface, whereas at the take-off angle of 90$\degree$, the emitted photoelectrons are from a deeper depth within the sample. Hence, the measurement gets more bulk sensitive as the photoelectron take-off angle increases from 35$\degree$ to 90$\degree$, and as the excitation source is changed from Al to Ag. The parameters of the fits pertaining to the angle-dependent analysis are given in Table~\ref{Tab1}.  The ratio of the Ba~II feature was observed to decrease considerably with bulk sensitivity measurements, thus confirming its surface character.

Two satellite features labeled Sat~I and Sat~II were also detected in the XPS spectra around the Ba~3d peaks [Fig.~\ref{Fig3}]. These satellites result from shake-up processes involving Ba~3d photoelectrons and valence electrons~\cite{Santana2016,Armen1985}.  These broad satellites are located about 10~eV on the high binding energy side of the associated  Ba~3d$_\frac{\text{5}}{\text{2}}$ and Ba~3d$_\frac{\text{3}}{\text{2}}$ peaks.

Figure~\ref{Fig4} depicts the Sn~3d core level XPS spectra together with the Voigt function fits after subtraction of a Shirley background, for three representative samples of  different total carrier density $n$. The Sn~3d spectral lineshapes display an asymmetry to the high binding energy side of the peak, which increases with increasing carrier density. In metallic systems, the asymmetry in core photoemission spectra arises from intrinsic plasmon excitations associated with the creation of the core hole, which results in an additional component satellite to the main core line \cite{Egdell1999a,Egdell2003a}. It is known that the Coulomb potential of the core hole creates a localized trap state by capturing a conduction electron \cite{Chazalviel1977a,Campagna1975a,Korber2010}. In La:BaSnO$_\text{3}$ systems, the conduction band is derived from highly dispersive Sn~5s bands \cite{Sallis2013c,Lebens-Higgins2016a}, and the observed doping effect in the Sn~3d core level  lineshape is most probably due to screening responses of the conduction electrons introduced by doping \cite{Egdell1999a}.  Therefore, two doublet components were used to fit the Sn~3d spectra assuming that the Koopmans' state (i.e. the excited state after the removal of a core electron from the atom) is projected into screened and unscreened final eigenstates \cite{Egdell1999a}.
\begin{figure}[!t]
	\centering 
	\includegraphics[width=0.45\textwidth]{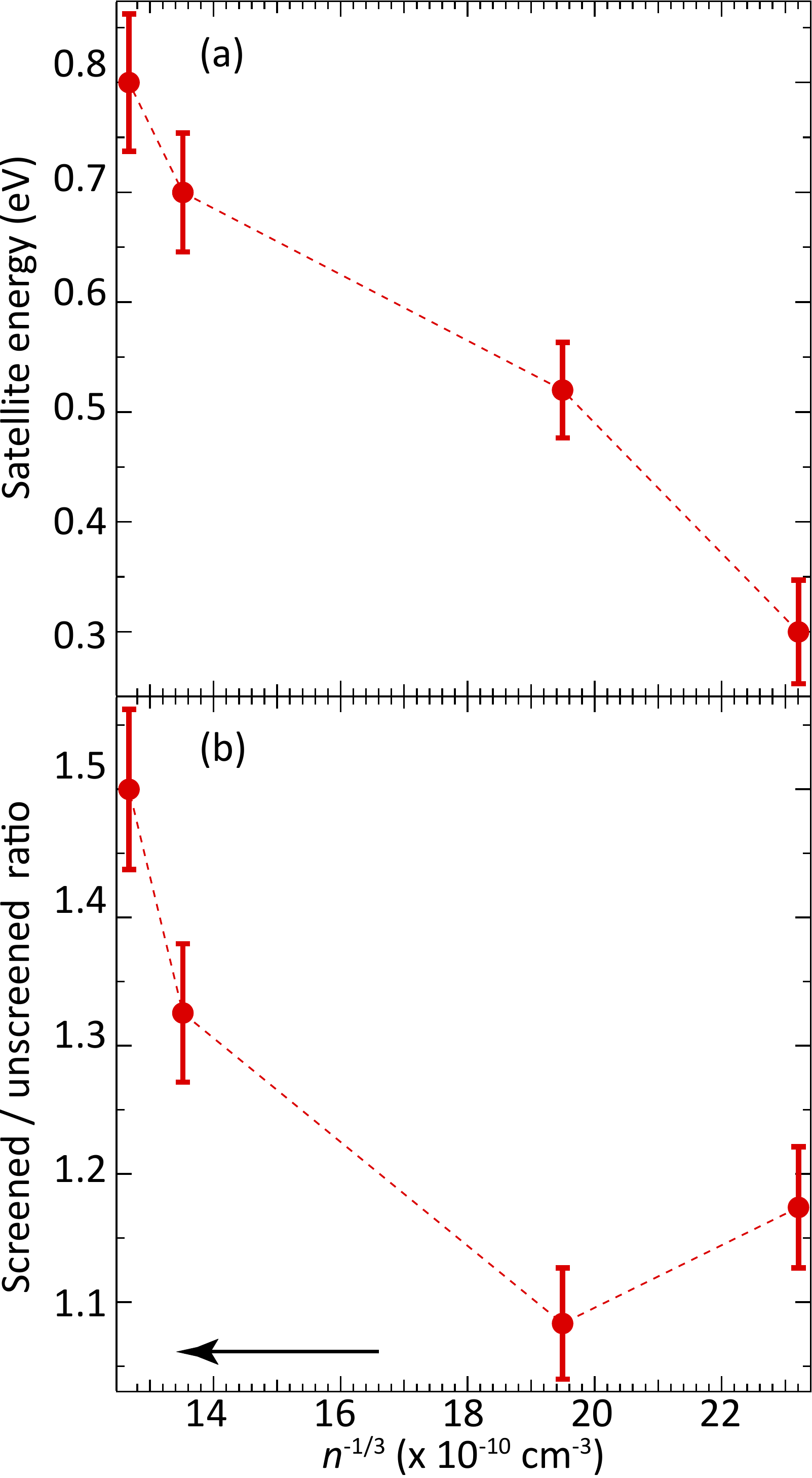}
	\caption{\label{Fig_bis}\small{Variation of the (a) satellite energy and (b) intensity ratio of the screened and unscreened components as a function of $n^{-1/3}$. The error bars in the carrier density are too small to be seen in this plot. The dashed lines are guides to the eye. The black arrow indicates increasing $n$ (plotted data are from Table~\ref{Tab2}).}}	 
\end{figure}

To understand the effect of increasing carrier density in the Sn~3d core level, the spectral lineshape of four samples (B, C, D and E) of different $n$ values were investigated and the analysis results are summarized in Table~\ref{Tab2}.  The core lines were fitted to two Voigt components, which give an excellent description of the overall line shape of the spectra. In each spectrum, the main component is the peak at low binding energy labeled ``screened''. This peak has a dominant Gaussian line shape. The component at high binding energy labeled ``unscreened" is dominantly Lorentzian, which is a satellite associated with intrinsic plasmon excitations \cite{Korber2010}. This peak is broader than the screened component, as evidenced by their FWHM values.  Similar satellite structures were reported in the Sn~3d and In~3d core photoemission spectra of binary transparent conducting oxides (Sb-doped SnO$_\text{2}$ \cite{Egdell2003a,Cox1982a,Egdell1999a}, In$_\text{2}$O$_\text{3}$--ZnO \cite{Jia2013a} and Sn-doped In$_\text{2}$O$_\text{3}$ \cite{Christou2000a,Korber2010}). 
\begin{figure*}[t]
	\centering 
	\includegraphics[width=0.95\textwidth]{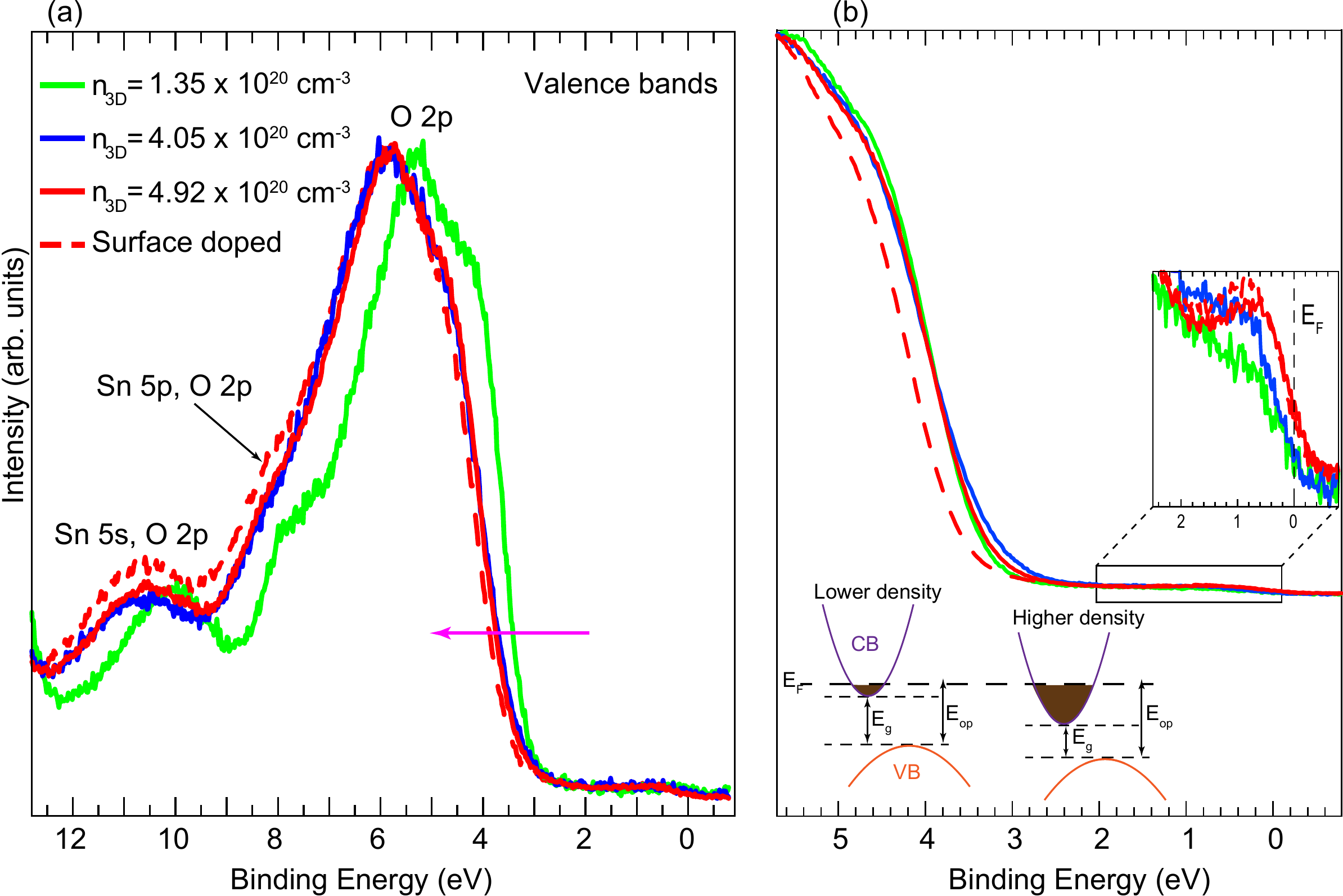}
	\caption{\label{Fig5}\small{(a) XPS valence band spectra of three samples of different total carrier density excited with the Al~K$_{\alpha}$ anode: Sample  D (green curves), Sample C (blue curves), and Sample B (red and red dashed curves). To investigate how surface absorbed carbonate and hydroxide layers affect the states in the valence band region, after initial measurements of the core levels and valence band spectra, the sample B (red dashed curve, surface doped) was intentionally exposed to contamination in the loadlock chamber  operated at 1$\times \text{10}^{-\text{8}}$~mbar (not UHV). The carrier densities of all the samples are indicated. The blue, green and red dashed curves are normalized to the maximum intensity of the red curve. The magenta arrow indicates the shift of the valence band with increasing carrier density. (b) Magnified view of the  low energy part of the spectra on (a). The bottom left inset is a schematic illustration  of the Moss-Burstein shift, which shows how the doping process affects the electronic band structure of the material. The top right inset is enlarged spectra for the region around the Fermi level.}}
\end{figure*} 
To better visualize the connection between carrier density  and the screened/unscreened intensity and energy, the evolution of both the satellite energy and the intensity of the peaks with $n^{-1/3}$ is plotted [Fig.~\ref{Fig_bis}]. As can be seen from Table~\ref{Tab2}, the binding energy value of the main component suggests a valence state of ${\text{4}+}$ for Sn \cite{Morgan1973,Cris2000}; and the energy separation (satellite energy) between the main and satellite components increases with $n$ [Fig.~\ref{Fig_bis}\textcolor{blue}{(a)}]. Furthermore, the relative intensity of the screened component increases with increasing $n$, while that of the unscreened peak decreases. This is indicated in Fig.~\ref{Fig_bis}\textcolor{blue}{(b)} by the increase of the intensity ratio of the peaks, conveying good agreement with previous reports \cite{Langreth1973a,Chazalviel1977a,Egdell2003a,Jia2013a,Korber2010}.  The discrepancy at the lowest $n$ (sample E) could be ascribed to the fact that the thickness of this sample, 100~nm, surpasses a critical thickness above which additional structural defects are induced in the film \cite{berger1988role,kneiss2020growth}. We speculate that defect scattering dominates, as only 15\% of the carriers are activated  compared to more than 50\% activation for the other samples. 

For fitting the spectra,  the Gauss-Lorentz ratio was allowed to vary freely. Boundaries were set for the satellite energy. These constraints were applied to the lower limit of the position of the unscreened peak with the consideration that the satellite energy corresponds to the plasmon energy (i.e., the surface plasmon energy) and increases with the carrier density  \cite{Egdell2003a,Egdell1999a,Korber2010,Payne2005,Glans2005}. Since the surface plasmon energy is proportional to the carrier density \cite{Egdell2003a,Christou2000a}, the constraints were such that  the satellite energy would be in the range of values reported for photoemission spectra of the 3d orbitals in binary transparent conducting oxides \cite{Egdell2003a,Egdell1999a,Jia2013a,Korber2010} with comparable $n$ values as in   samples B, C, D and E. Therefore, the consistent observation of a narrower low binding energy peak and a broader high binding energy peak supports the applicability of the plasmon model to the analysis of the Sn~3d core XPS spectra in these films, which is consistent with previously reported Sn~3d core level spectra in Sb-doped SnO$_\text{2}$ samples~\cite{Egdell2003a,Cox1982a,Egdell1999a}.

Next, the effect of increasing carrier density on the valence and conduction band spectra was explored. The same La:BaSnO$_\text{3}$ samples characterized for the core level spectra were used. The XPS spectra of the valence  and conduction bands are depicted in Figs.~\ref{Fig5}\textcolor{blue}{(a)} and \ref{Fig5}\textcolor{blue}{(b)}, respectively. 

 Three main features are observed in the valence band spectra: (i) a mixture of Sn~5s and bonding O~2p orbitals located at 10.6~eV; (ii) the states at 8.3~eV originating from hybridized Sn~5p and O~2p orbitals; and (iii)  the bands at binding energies between 4 and 6~eV associated with O~2p bonding or anti-bonding character \cite{Themlin1990,Farahani2014,Kover1995}. Additionally,   shifts of the valence band leading edge to high binding energies upon increasing carrier density can be observed. This indicates an increase in the optical band gap as proposed previously in  ellipsometry and HAXPES results~\cite{Lebens-Higgins2016a,Niedermeier2017d,Seo2014a}. These shifts are correlated to the shifts observed in the core levels [See section \textcolor{blue}{S2} of the supplementary information], as well as to the increasing asymmetry in the Sn~3d core lines. Similar trends were observed in other degenerate doped transparent conducting oxides, which were attributed to the increasing occupation of the states in the conduction band~\cite{Egdell1999a,Jia2013a,Korber2010}. It  is  noteworthy  that an opposite trend (i.e., shift of the valence band spectra toward lower binding energies with increases in La doping) was reported in recent angle resolved photoemission spectroscopy (ARPES) experiments on La:BaSnO$_3$ films, and it was suggested to originate  from the opposite evolution of surface and bulk chemical potentials~\cite{Lochocki2018b}.

\begin{figure}[!t]
	\centering 
	\includegraphics[width=0.47\textwidth]{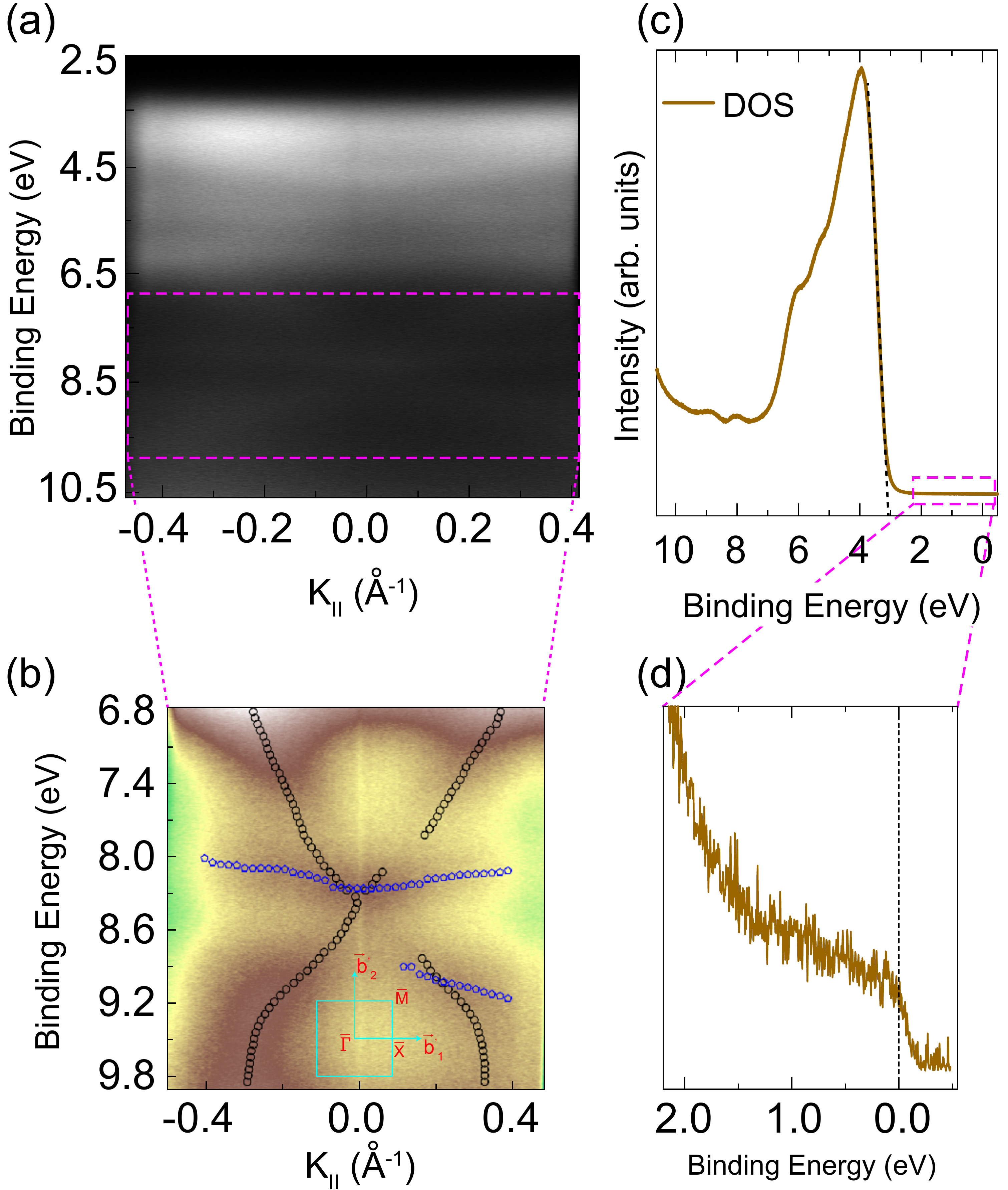}
	\caption{\label{Fig6}\small{(a) Representative 2D ARPES map for sample D. The map was acquired in the $\bar{\Gamma}-\bar{\text{X}}$ direction. (b)  ARPES map of the region indicated by the pink dashed rectangle. The inset shows the first Brillouin zone together with the $\bar{\Gamma},\, \bar{\text{X}} \text{ and } \bar{\text{M}}$ high symmetry points. (c)  A representative density of states (DOS) integrated over the entire momentum space from the ARPES map in (a). The black dashed line is a linear extrapolation of the valence band leading edge, revealing that the valence band maximum is located at $\sim 3.1$ eV. (d) Enlargement of the region around the Fermi level in (c).}}
\end{figure} 

To explore further spectral features arising from occupation of the conduction bands, high resolution scans around the valence band leading edges were acquired [Fig.~\ref{Fig5}\textcolor{blue}{(b)}]. To achieve an adequate signal to noise ratio and resolve the fine features in the region close to the Fermi energy (E$_\text{F}$), each spectrum was acquired over a period of about 20 hours.  A peak at $\sim$ 4~eV deriving from O~2p orbitals is observed in all spectra as indicated by their first derivatives [see Fig.~\textcolor{blue}{S5} of the supplementary  information]~\cite{Lochocki2018b,Joo2017b}. For the contaminated surface (sample with the highest $n$ intentionally exposed to contamination in the load lock), a shift  to higher binding energy of $\sim$ 0.23~eV is clearly visible in the leading edge valence band spectrum [see red dashed curve in Fig.~\ref{Fig5}\textcolor{blue}{(b)}]. Furthermore, a bump is detectable in all spectra in the region between 2 eV and E$_\text{F}$ [see top right inset in Fig.~\ref{Fig5}\textcolor{blue}{(b)}]. The spectra exhibit a weak structure close to E$_\text{F}$, which terminates in a sharp Fermi edge. This structure is associated with occupied states in the conduction band (Sn~5s orbital character with a small contribution from O~2p orbitals) \cite{Sallis2013c,Lebens-Higgins2016a}. Moreover, the intensity of this CBM peak is observed to increase in the contaminated surface as  evidenced by the red dashed curve [see top right inset in Fig.~\ref{Fig5}\textcolor{blue}{(b)}]. This suggests that exposure of the surface to contamination results in  increasing occupied states in the conduction band. We attribute this behavior to the Moss-Burstein effect, i.e.,  the apparent optical band gap of the material is increased as the absorption edge is pushed to higher energies as a result of some states close to the conduction band being populated [see, bottom left inset in Fig.~\ref{Fig5}\textcolor{blue}{(b)}]~\cite{EBurstein_1954,TSMoss_1954}. Indeed, the valence band shifts associated with the increasing density of electrons occupying the conduction band were reported in several transparent conducting oxides, resulting in the increase of the intensity of the conduction band feature~\cite{Korber2010,Lebens-Higgins2016a,Jia2013a,JJMudd_2014,Niedermeier2017d}. 

To date, few ARPES studies of the electronic band structure of La:BaSnO$_3$ and BaSnO$_3$ films have been reported~\cite{Lochocki2018b,Joo2017b,Soltani2020}. In Fig.~\ref{Fig6}, we present the ARPES data of the band structure of a representative La:BaSnO$_3$ film (Sample D) exposed to air for days before cleaning in vacuum~\cite{ARPES_2019}. These data were collected at  room temperature. Although ARPES is a very surface sensitive technique and the samples were exposed to ambient conditions, valence band dispersion is observed from about 3.10 to 10.62~eV [Fig.~\ref{Fig6}\textcolor{blue}{(a)}]. The fact that these bands are not clearly resolved is understood in terms of the need for a very particular surface treatment associated with \textit{ex-situ} ARPES measurements~\cite{Soltani2020}.  Figure~\ref{Fig6}\textcolor{blue}{(b)} depicts a high-resolution 2D ARPES map in the binding energies ranging from 6.8 to $\sim$ 9.8~eV around the  hybridized Sn~5s and O~2p states. Some highly dispersive bands are resolved: the black markers overlaying the dispersing bands are extracted band dispersions obtained from peak fitting of the momentum distribution curves (MDCs), whereas the blue markers are for bands fitted to peaks in the energy distribution curves (EDCs). Figure~\ref{Fig6}\textcolor{blue}{(c)} represents the density of states (DOS) integrated from the ARPES map (i.e., EDC obtained from the ARPES map over the entire momentum space). The DOS spectrum exhibits well resolved band features that are similar to the XPS valence band spectra [Fig.~\ref{Fig5}\textcolor{blue}{(a)}]. The  Fermi-Dirac edge straddling 0~eV is visible [Fig.~\ref{Fig6}\textcolor{blue}{(d)}], and a linear extrapolation of the valence band leading edge reveals that the VBM is situated at $\sim 3.1$ eV [see, black dashed line in Fig.~\ref{Fig6}\textcolor{blue}{(c)}]. The extracted VBM value is in close agreement to previous theoretical and experimental values~\cite{Sallis2013c,Lochocki2018b,Joo2017b,Soltani2020}.  For this sample investigated in both XPS and ARPES, the same value for the VBM is obtained from both techniques; and the CBM is not well developed as evidenced by the data for both techniques shown in the insets in Fig.~\ref{Fig5}\textcolor{blue}{b} and Fig.~\ref{Fig6}\textcolor{blue}{d}.

In summary, we have systematically investigated the evolution of electronic states in the band structure of La:BaSnO$_\text{3}$ films at different La doping levels. A close connection between the transport and the  spectroscopic characteristics is demonstrated. In particular, increasing the carrier concentration in the conduction band by doping is observed to significantly affect the core and valence band spectra.  The Sn~3d core line shape  presents a pronounced asymmetry variation with the carrier density, and is fitted following the plasmon model applicable to metallic systems.  Scans around the valence band spectra allowed the detection of the occupied states in the conduction bands. It is determined that surface contamination could potentially induce surface carrier accumulation, supported by the increase in the intensity of the CBM detected in the surface exposed to contamination. This study presents a detailed characterization of the chemical composition of the near-surface region of La:BaSnO$_\text{3}$, and it provides a  better  picture of the interplay between the doping concentration, electronic band structure and transport properties of epitaxial La:BaSnO$_\text{3}$ films. The ARPES data presented in this study highlight the challenge of surface preparation for ex-situ ARPES measurements of epitaxial La:BaSnO$_\text{3}$ films and heterostructures. Preferably, a portable vacuum suitcase is ideal for long-distance transport of epitaxial La:BaSnO$_\text{3}$ films in UHV conditions to perform further in situ ARPES analysis at different locations~\cite{PNgabonziza_situ_2018} and/or development of protective capping methods for in situ capping to avoid any possible surface contamination when films are exposed to ambient conditions.
\\ \\
\textbf{Supplementary Material}\\
\indent See supplementary material for details on electronic transport measurements, additional microstructural and photoemission spectroscopy characterizations of La:BaSnO$_3$ films and heterostructures. 
\\ \\
\indent 
The authors grateful acknowledge fruitful discussions and  technical support by Kathrin Küster. B. P. Doyle, A. P. Nono Tchiomo, A. R. E. Prinsloo and  E. Carleschi acknowledge funding support from the National Research Foundation (NRF) of South Africa under Grant Nos. 93205, 90698, 99030, and 111985. W. Sigle and P. van Aken acknowledge funding from the European Union’s Horizon 2020 research and innovation programme under grant agreement No. 823717 -ESTEEM3.
\\ \\
\textbf{REFERENCES}
\bibliography{references-2021}

\begin{thebibliography}{86}
\bibitem [{ARP()}]{ARPES_2019}%
  \BibitemOpen
  \href@noop {} {\bibinfo  {journal} {The La:BaSnO$_3$ films were prepared at
  the Max Planck Institute for Solid State Research, Stuttgart, Germany and
  transported at ambient conditions for further spectroscopic measurements (XPS
  and ARPES) to the University of Johannesburg, South Africa. Details on the
  cleaning procedure of the samples and spectroscopic measurements are
  presented in supplemantry information}\ }\BibitemShut {NoStop}%
\end{thebibliography}
\onecolumngrid
\newpage
\setcounter{table}{0}
\setcounter{figure}{0}
\renewcommand{\thefigure}{S\arabic{figure}}%
\setcounter{equation}{0}
\renewcommand{\theequation}{S\arabic{equation}}%
\setstretch{1.5}
\begin{center}
\title*{\textbf{\Large{Supplementary Information:} \\ [0.25in] \large{Combined Spectroscopy and Electrical Characterization of La:BaSnO$_\text{3}$ \\Thin Films and Heterostructures}}}
\end{center}

\begin{center}
\large{Arnaud P. Nono Tchiomo,$^{1,\,2}$ Emanuela Carleschi,$^1$ Aletta R. E. Prinsloo,$^1$ Wilfried Sigle,$^3$ Peter A. van Aken,$^3$ Jochen Mannhart,$^3$ Prosper Ngabonziza,$^{1,\,3,{\textcolor{blue}{\small{\,a)}}}}$ Bryan P. Doyle,$^{1,{\textcolor{blue}{\small{\,b)}}}}$}\newline 
\large{\textit{$^1$Department of Physics, University of Johannesburg, P.O. Box 524,  Auckland Park 2006,  Johannesburg, South Africa
\newline $^2$Van der Waals-Zeeman Institute, Institute of Physics, Science Park 904, 1098 XH Amsterdam, The Netherlands \newline $^3$Max Planck Institute for Solid State Research, Heisenbergstr. 1, 70569 Stuttgart, Germany}}
\end{center}

\section*{\large{S1:} \large{E\lowercase{lectronic} T\lowercase{ransport} M\lowercase{easurements and }M\lowercase{icrostructural} C\lowercase{haracterizations} }}

\begin{figure*}[!t]
\centering
     \includegraphics[width=0.72\textwidth]{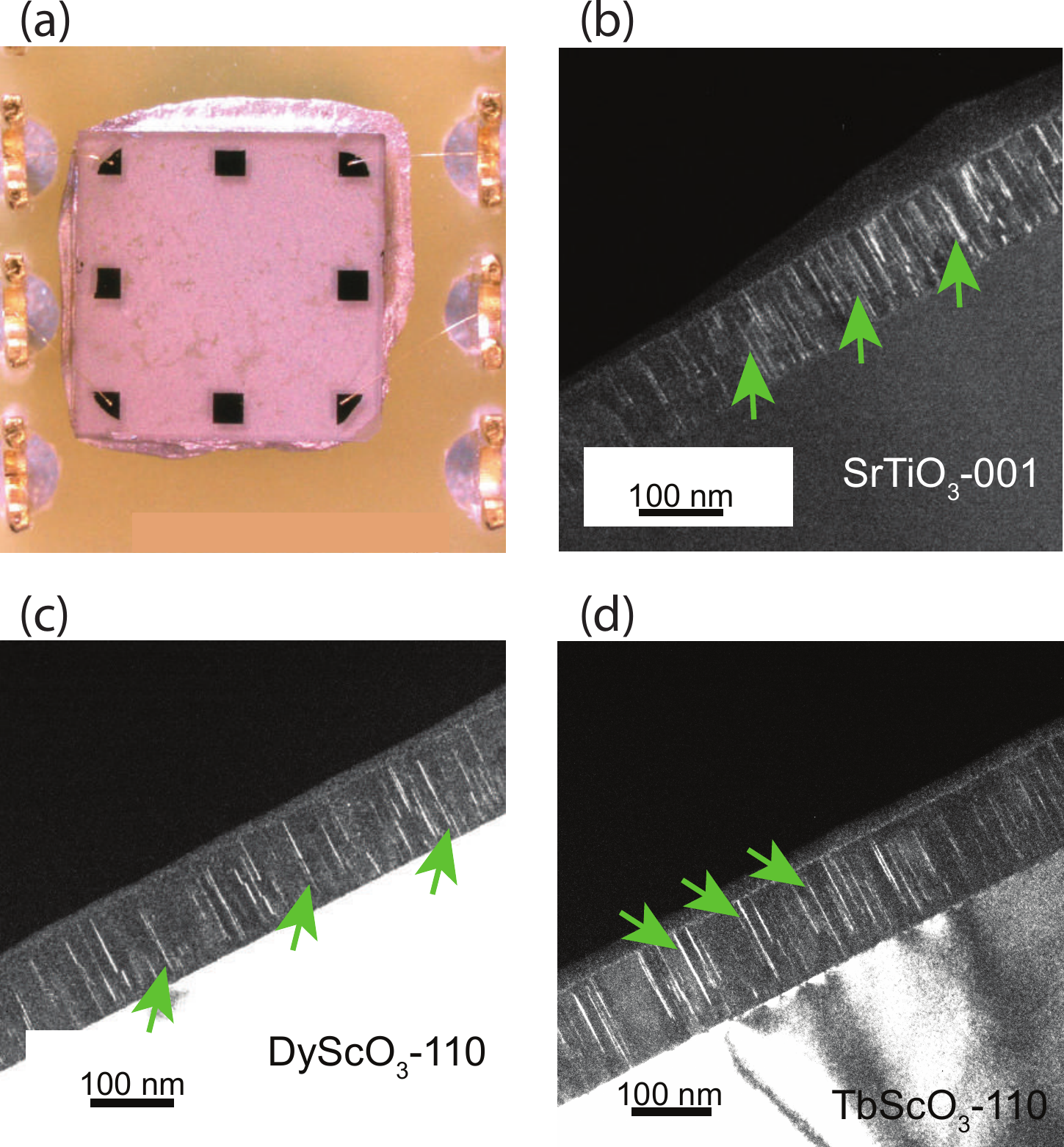}
    \caption{\small{(a) Optical photograph of a typical La:BaSnO$_3$ film structured in the Van der Pauw geometry, using metal contact pads of  Au/Ti. The contacts are connected to the sample holder pins using aluminum wires bonded at the corners of the sample. (b) - (d) WB-DFTEM micrographs of (b) BaSnO$_3$/SrTiO$_3$, (c) BaSnO$_3$/DyScO$_3$ and (d) BaSnO$_3$/TbScO$_3$ heterostructures, showing threading dislocations (vertical bright contrasts) running across the film from the interface and represented by green arrows.}} 
\label{FigS1}
\end{figure*}

All La:BaSnO$_3$ samples were grown by pulsed laser deposition at an  optimal target-substrate distance of 56 mm. 
The deposition parameters are given in Table~\ref{table_growth}. After growth of La:BaSnO$_3$ films, Au/Ti contacts (45 nm thick) that provide electrical connections to the sample were added to the  samples'  corners using standard photolithography methods [Fig.~\ref{FigS1}\textcolor{blue}{(a)}]. Basic electrical characterizations of the samples were performed using a physical property measurement system (PPMS) in four-point configuration with an excitation  current  of  1$\mu$A. 

\begin{table*}[!b]
	\begin{center}
		\caption{\small{Growth parameters of the La:BaSnO$_\text{3}$ thin films  discussed in this study.}}
		\label{table_growth}
		\begin{tabular}{ccccccc}
			\hline
			\hline
			\small{Sample} & \multicolumn{1}{c}{\small{Thin film}}  & \small{Substrate} & \small{Deposition} & \small{O$_\text{2}$ partial} & \small{Laser} & \small{Laser repetition} \\
			& \multicolumn{1}{c}{\small{layer}} &  &  \small{temperature} & \small{pressure} & \small{fluence} & \small{rate} \\
			& &  &  \small{($\degree$C)} & \small{(mbar)} &\small{(J~$\cdot$~cm$^{-\text{2}}$)} & \small{(Hz)}\\
			\multicolumn{1}{c}{} & \multicolumn{1}{c}{\small{BaSnO$_\text{3}$ (100~nm)}} & \small{} & \small{850} & \small{$\text{1.00}\times\text{10}^{-\text{1}}$} & \small{1.5} & \small{4} \\
			\multicolumn{1}{c}{\multirow{-2}{*}{\small{A}}} & \multicolumn{1}{c}{\small{6\% La:BaSnO$_\text{3}$ (25~nm)}} &\multicolumn{1}{c}{\multirow{-2}{*}{\small{DyScO$_\text{3}$}}} & \small{850} & \small{$\text{1.00}\times\text{10}^{-\text{1}}$} & \small{1.5} & \small{1} \\
			\multicolumn{1}{c}{\multirow{2}{*}{\small{B}}} & \multicolumn{1}{c}{\small{BaSnO$_\text{3}$ (100~nm)}} & \multicolumn{1}{c}{\multirow{2}{*}{\small{TbScO$_\text{3}$}}} &\small{850} & \small{$\text{1.00}\times\text{10}^{-\text{1}}$} & \small{1.5} & \small{5} \\
			\multicolumn{1}{c}{} & \multicolumn{1}{c}{\small{6\% La:BaSnO$_\text{3}$ (25~nm)}} & \multicolumn{1}{c}{} & \small{850} & \small{$\text{1.00}\times\text{10}^{-\text{1}}$} & \small{1.5} & \small{5} \\
			\multicolumn{1}{c}{\small{C}} & \multicolumn{1}{c}{\small{6\% La:BaSnO$_\text{3}$ (25~nm)}} & \small{SrTiO$_\text{3}$} & \small{850} & \small{$\text{1.05}\times\text{10}^{-\text{1}}$} & \small{1.5} & \small{1} \\
			\multicolumn{1}{c}{\small{D}} & \multicolumn{1}{c}{\small{2\% La:BaSnO$_\text{3}$ (25~nm)}} & \small{TbScO$_\text{3}$} & \small{850} & \small{$\text{1.05}\times\text{10}^{-\text{1}}$} & \small{1.5} & \small{1} \\
			\multicolumn{1}{c}{} & \multicolumn{1}{c}{\small{BaSnO$_\text{3}$ (25~nm)}} & \small{} & \small{850} & \small{$\text{1.33}\times\text{10}^{-\text{1}}$} & \small{1.5} & \small{1} \\
			\multicolumn{1}{c}{\multirow{-2}{*}{\small{E}}} & \multicolumn{1}{c}{\small{4\% La:BaSnO$_\text{3}$ (100~nm)}} &\multicolumn{1}{c}{\multirow{-2}{*}{\small{SrTiO$_\text{3}$}}} & \small{850} & \small{$\text{1.33}\times\text{10}^{-\text{1}}$} & \small{1.5} & \small{1} \\
			\hline
			\hline
		\end{tabular}
	\end{center} 
\end{table*}

The scanning transmission electron microscopy (STEM) and electron energy loss spectroscopy (EELS)  investigations were performed using a Cs-probe-corrected JEOL JEM-ARM200F. The EELS measurements were performed on the cross-section of BaSnO$_\text{3}$ buffered samples, and the electron beam was focused on the  buffer layer and the  La:BaSnO$_\text{3}$ film layer regions. Spectra from areas without and with La could then be compared, to see whether the doping alters or changes the oxidation state of the Sn. 

In epitaxial thin film growth, a buffer layer can be employed to boost both the structural and electronic properties of the film \textcolor{blue}{[1, 2, 3]} . The primary role of this additional layer is to reduce the density of defects, and in particular, the concentration of threading dislocations, which result in carrier density and carrier mobility reduction \textcolor{blue}{[1, 2, 3]}. These dislocations are known to originate from misfit dislocations, which are generated at the film/substrate interface due to the large lattice mismatch between the as-deposited film layer and the substrate \textcolor{blue}{[4, 5, 6]}. Figures~\ref{FigS1}\textcolor{blue}{(b)}-\textcolor{blue}{(d)} depict scanning transmission electron microscopy (STEM)  micrographs of BaSnO$_3$/SrTiO$_3$, BaSnO$_3$/DyScO$_3$ and BaSnO$_3$/TbScO$_3$ heterostructures. Dense edge-type threading dislocations identified as vertical lines running across the film from the interface can be observed \textcolor{blue}{[4, 6]}.  A quantitative analysis of these threading dislocations carried out by counting the number of lines in 500~nm by 16~nm thick specimens gives a dislocation density of $\text{10}\times \text{10}^{\text{11}}$~cm$^{-\text{2}}$ in the film prepared on SrTiO$_3$, and $\text{5}\times \text{10}^{\text{11}}$~cm$^{-\text{2}}$ in the films grown with DyScO$_3$ and TbScO$_3$ substrates. This high concentration of threading dislocations acts as charge traps, and was reported to be responsible for the low activation rate of the  carrier, as well as for the limitation of the electron  mobility in La:doped BaSnO$_3$ systems \textcolor{blue}{[1, 7, 8]}.

\section*{\large{S2:} \large{P\lowercase{hotoemission} S\lowercase{pectroscopy} M\lowercase{easurements}}}
\begin{figure*}[!t]
	\centering 
	\includegraphics[width=0.57\textwidth]{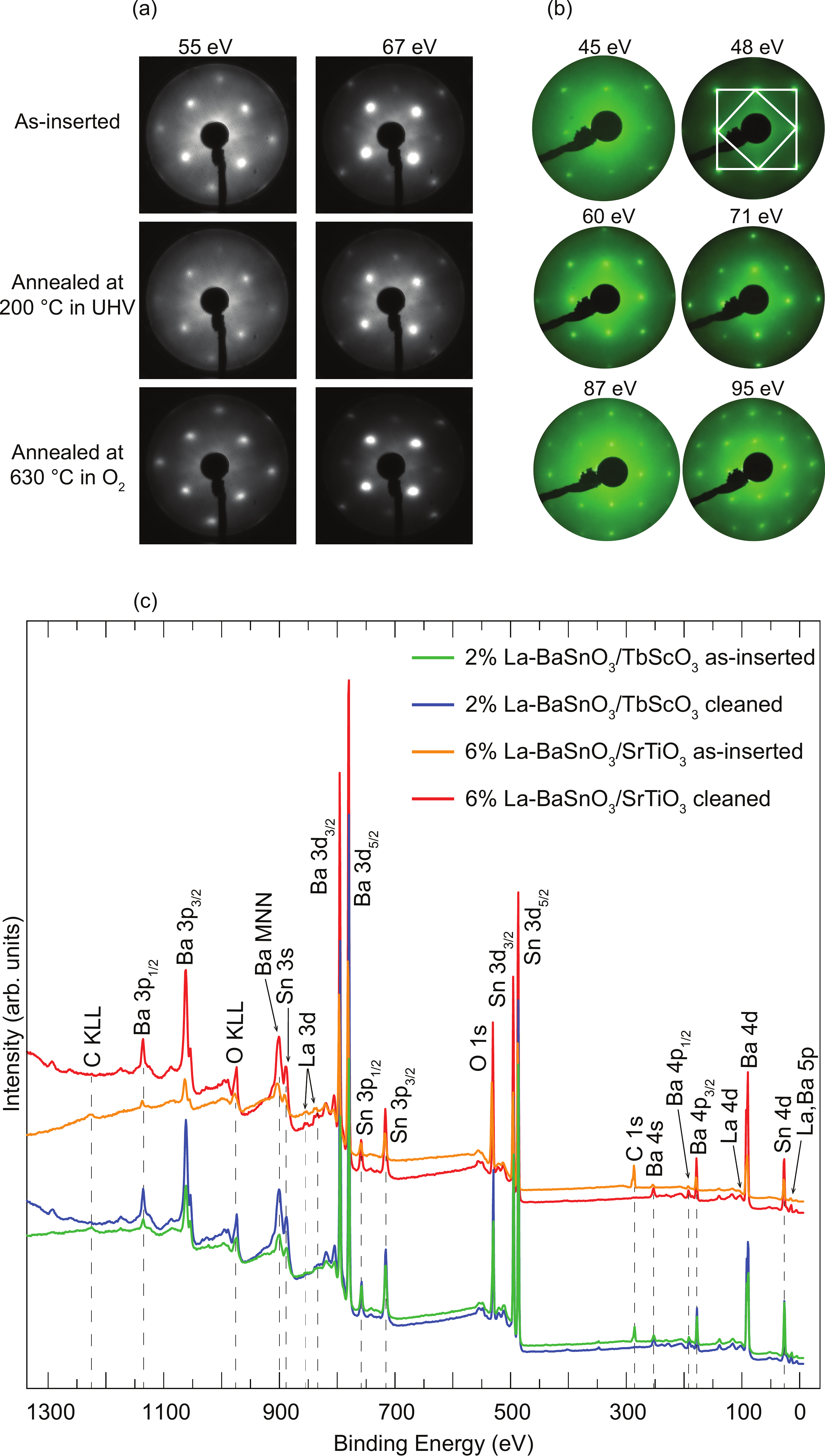}
	\caption{\label{FigS2}\small{(a) and (b) Representative LEED images from the surface of a La:BaSnO$_3$  (25 nm)/BaSnO$_3$ (100 nm)/TbScO$_3$ heterostructure. (a) The images are taken at various electron energies on the surface at different stages of cleaning. (b) The images are taken at various electron energies on the clean surface. The white squares drawn on the image recorded at 48 eV illustrate the square lattice of BaSnO$_3$. Note that at the electron energy of 48 eV, the diffraction pattern shows only first order beams, and when the energy is approximately doubled (at 95 eV), second order spots are observed. (c) XPS survey scans for as-inserted (light green and orange curves) and cleaned (blue and red curves) for the samples E and D. All the core levels from La, Ba, Sn and O elements can be seen in the spectra. The C 1s signal originating from surface contamination with the associated C KLL Auger electron peak are also visible. The spectra were acquired with an experimental resolution of 0.90 eV using the Al K$_{\alpha}$ excitation source.}}	 
\end{figure*}

The La:BaSnO$_3$ thin films and heterostructures were grown using pulsed laser deposition at the Max Planck Institute for Solid State Research, Stuttgart Germany. After electronic transport characterizations, samples were transported in ambient conditions to the University of Johannesburg in South Africa for further spectroscopic measurements. X-ray photoemission spectroscopy (XPS) and angle resolved photoemission spectroscopy (ARPES) spectra were collected using a SPECS PHOIBOS 150 hemispherical electron energy analyzer. The base pressure in the  analysis chamber during acquisition of the core and valence XPS data was $\approx$~3$\times \text{10}^{-\text{10}}$~mbar. These were measured using  monochromatized Al~K$_{\alpha}$ and Ag~L$_\alpha$ excitation sources, emitting photons of energies  of 1486.71~eV and 2984.31~eV, respectively. The experimental resolutions with these two sources were 580 and 960~meV, respectively. To investigate the electronic band structure, we also excited the valence electrons of the samples using a He~I source delivering photons of 21.2~eV. For these ARPES measurements, the experimental resolution was 120~meV.
\begin{figure*}[!b]
	\centering 
	\includegraphics[width=1\textwidth]{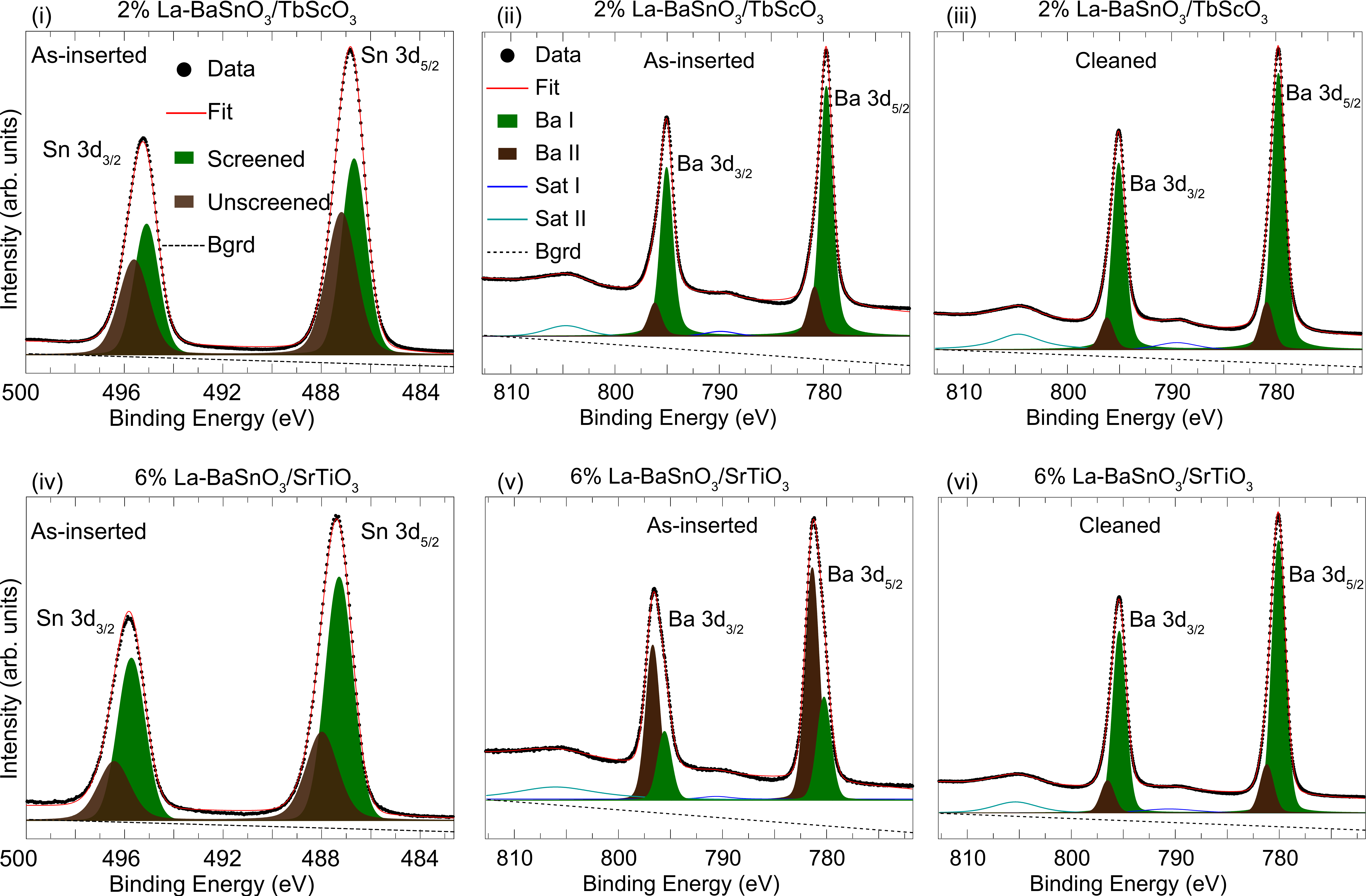}
	\caption{\label{FigS3}\small{Fits of the XPS spectra around the Sn~3d and Ba~3d regions for as-inserted and cleaned samples D (i-iii) and  C (iv-vi), discussed in the main text.  The spectra were fitted with two Voigt doublets. The data were acquired with an experimental resolution of 0.78 eV}}	 
\end{figure*}

\begin{figure*}[!h]
	\centering 
	\includegraphics[width=0.9\textwidth]{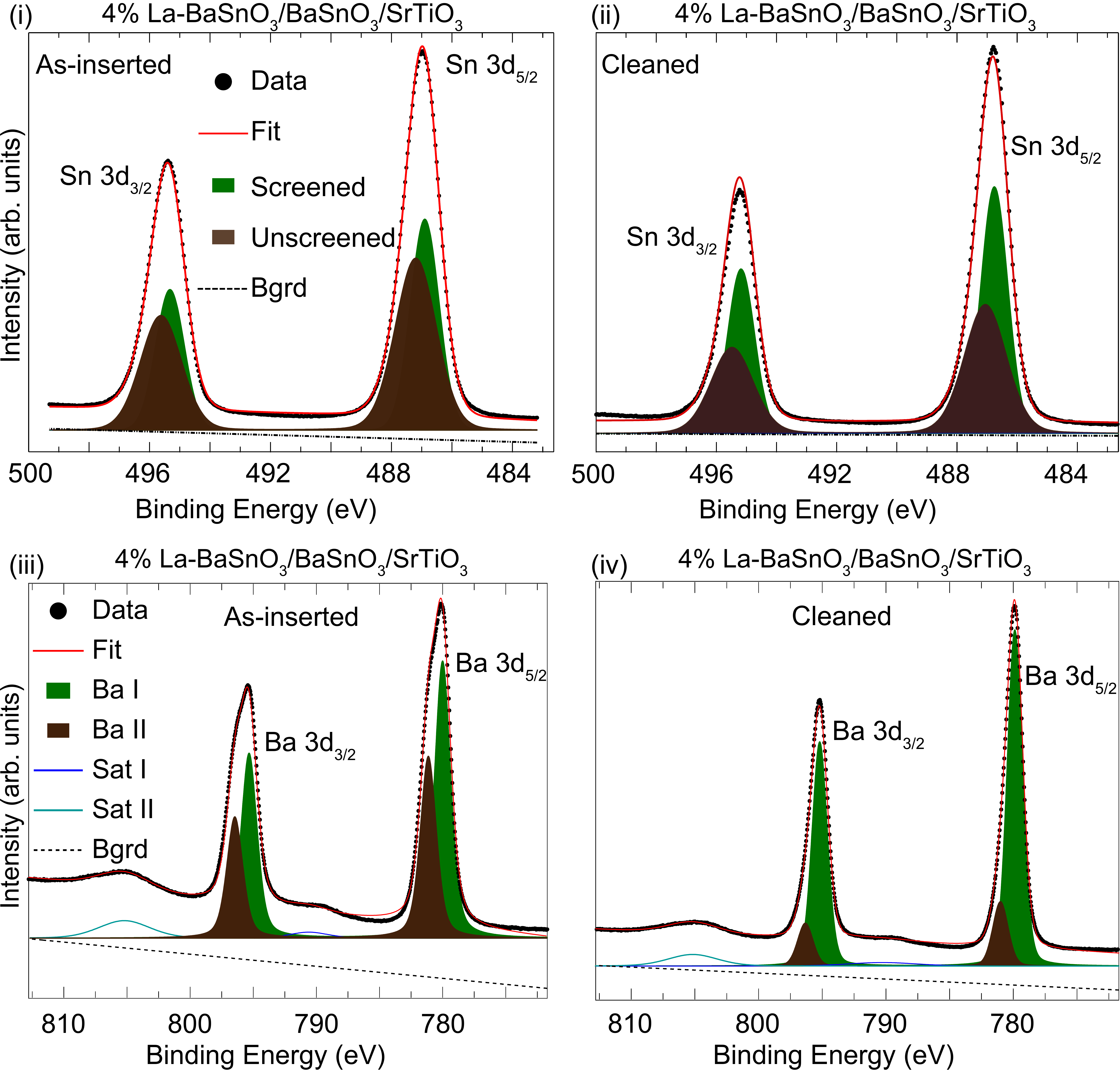}
	\caption{\label{FigS4}\small{Fits of the as-inserted and cleaned Sn~3d and Ba~3d XPS spectra for  sample E.}}	 \end{figure*}
\begin{figure*}[!t]
	\centering 
	\includegraphics[width=0.85\textwidth]{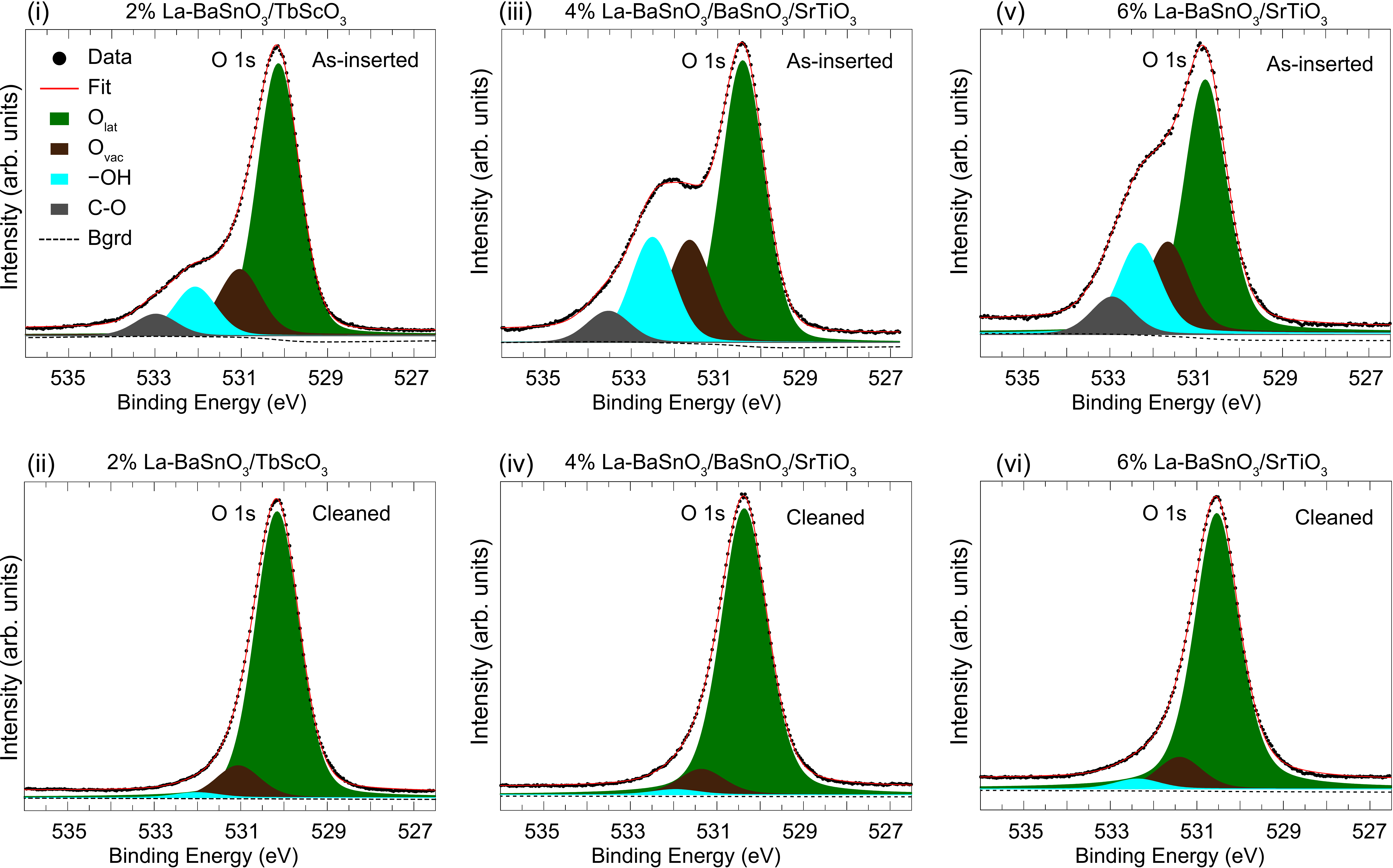}
	\caption{\label{FigS5}\small{Fits of the XPS spectra for the O~1s core levels for samples D (i and ii), E (iii and iv) and C (v and vi). The O 1s core electrons were fitted with four Voigt singlet components before cleaning, whereas three were used after vacuum cleaning steps. O$_{\text{lat}}$ is the main component attributed to lattice oxygen. The peak labeled O$_{\text{vac}}$ is a surface oxygen vacancy. This peak represents the change in the local electronic density of the other lattice oxygen atoms when an oxygen is removed \textcolor{blue}{[9, 10, 11]}. $-$OH and C-O are contamination related components, associated with adsorbed oxygen resulting from the dissociation of water on the surface \textcolor{blue}{[12, 13, 14]}, and oxygen attached to organic residues or carbonates \textcolor{blue}{[13, 14]}, respectively.}}	 
\end{figure*}
\begin{figure*}[!h]
	\centering 
	\includegraphics[width=0.8\textwidth]{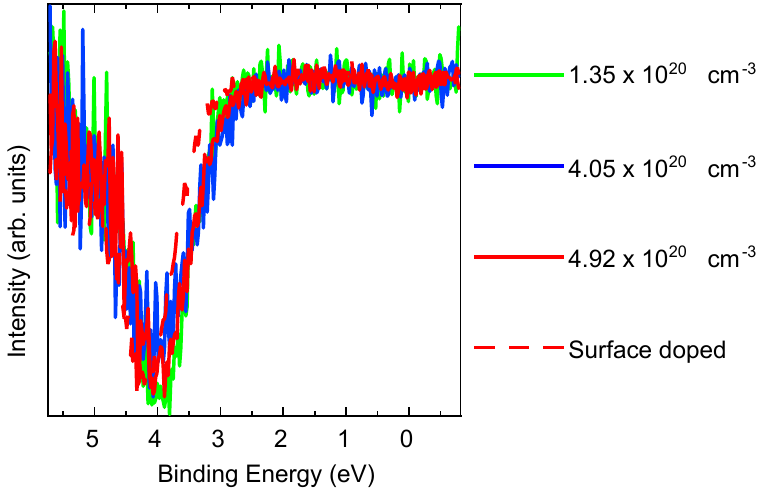}
	\caption{\label{FigS6}\small{First derivative of the XPS spectra discussed in Fig.~\textcolor{blue}{5} of the main text.}}	 
\end{figure*}

As the samples had been exposed to air under ambient conditions, their surfaces were thoroughly cleaned in vacuum prior to spectroscopic measurements. The cleaning procedure consisted of annealing in O$_\text{2}$ environment at the temperature of $\sim 700$~$\degree$C in a preparation chamber maintained at $\approx$~7$\times \text{10}^{-\text{6}}$~mbar. The annealing procedure consisted of several cycles of 2 hours each and it was done using an e-beam heater at a maximum electron emission power of 3.5~W. The temperature was indicated by a pyrometer set to an emissivity of 0.1. The cleanliness of the samples was monitored by recording the low-energy electron diffraction (LEED) patterns (to inspect the surface structure) directly after an annealing cycle [Fig.~\ref{FigS2}\textcolor{blue}{(a)}], as well as by tracking the C~1s signal in the XPS survey scans [Fig.~\ref{FigS2}\textcolor{blue}{(c)}] and by tracking the O~1s, Sn~3d and Ba~3d levels before and after surface cleaning [see, Fig.~\ref{FigS3}, Fig.~\ref{FigS5}]. LEED was also used for the determination of the orientation of the surfaces before ARPES measurements. Figure~\ref{FigS2}\textcolor{blue}{(a)} and Fig.~\ref{FigS2}\textcolor{blue}{(b)} depict representative LEED images of the La:BaSnO$_3$ $(001)$ surface at different stages in the high-temperature cleaning process, and at different electron energies (clean surface), respectively. 
\newpage
\begin{center}
\section*{\large{R\lowercase{eferences}}}
\end{center}

[1] A. P. Nono Tchiomo, W. Braun, B. P. Doyle, W. Sigle, P. van Aken, \newline \indent J. Mannhart, and P. Ngabonziza, APL Mater. \textbf{7}, 041119 (2019).

[2] S. Nakamura, Jpn. J. Appl. Phys. \textbf{30}, L1705 (1991).

[3] Y. Huang, X. D. Chen, S. Fung, C. D. Beling, C. C. Ling, Z. F. Wei, S. J. Xu, and C.
Y. Zhi, \newline \indent J. Appl. Phys. \textbf{96}, 1120 (2004).

[4] H. Paik, Z. Chen, E. Lochocki, H. Ariel Seidner, A. Verma, N. Tanen, J. Park, M. Uchida, \newline \indent S. Shang, B. C. Zhou, M. Brützam, R. Uecker, Z. K. Liu, D. Jena, K. M. Shen, D. A. Muller, and \newline \indent D. G. Schlom, APL Mater. \textbf{5}, 116107 (2017).

[5] S. Raghavan, T. Schumann, H. Kim, J. Y. Zhang, T. A. Cain, and S. Stemmer, \newline \indent APL Mater. \textbf{4}, 016106 (2016).

[6] J. Shiogai, K. Nishihara, K. Sato, and A. Tsukazaki, AIP Adv. \textbf{6}, 065305 (2016).

[7] Z. Lebens-Higgins, D. O. Scanlon, H. Paik, S. Sallis, Y. Nie, M. Uchida, N. F. Quackenbush,
 \newline \indent M. J. Wahila, G. E. Sterbinsky, D. A. Arena, J. C. Woicik, D. G. Schlom, and L. F. Piper,  \newline \indent Phys. Rev. Lett. \textbf{116}, 027602 (2016).

[8] H. J. Kim, U. Kim, T. H. Kim, J. Kim, H. M. Kim, B. G. Jeon, W. J. Lee, H. S. Mun,  \newline \indent K. T. Hong, J. Yu, K. Char, and K. H. Kim, Phys. Rev. B \textbf{86}, 165205 (2012).

[9] M. Sirena, N. Haberkorn, M. Granada, L. B. Steren, and J. Guimpel, \newline \indent J. Appl. Phys. \textbf{105}, 033902 (2009).

[10] J. Rubio-Zuazo, L. Onandia, P. Ferrer, and G. R. Castro,  \newline \indent Appl. Phys. Lett. \textbf{104}, 021604 (2014).

[11] K. Wang, Y. Ma, and K. Betzler, Phys. Rev. B \textbf{76}, 144431 (2007).

[12] A. Fujishima, X. Zhang, and D. A. Tryk, Surf. Sci. Rep. \textbf{63}, 515 (2008).

[13] J. C. Yu, J. Yu, H. Y. Tanga, and L. Zhang, J. Mater. Chem. \textbf{12}, 81 (2002).

[14] J. Zhuang, S. Weng, W. Dai, P. Liu, and Q. Liu, J. Phys. Chem. C \textbf{116}, 25354 (2012).

[15] G. Greczynski, and L. Hultman, Prog. Mater. Sci. \textbf{107}, 100591 (2020).

[16] E. Korin, N. Froumin, and S. Cohen, ACS Biomater. Sci. Eng. \textbf{3}, 882 (2017).

[17] D. Cabrera-German, G. Gomez-Sosa, and A. Herrera-Gomez,  \newline \indent Surf. Interface Anal. \textbf{48}, 252 (2016).
\end{document}